\documentclass[12pt]{article}
\usepackage{amsmath,amssymb,tikz-cd}
\usepackage{hyperref}
\usepackage{cite}
\usepackage{tikz}
\usetikzlibrary{arrows,chains,matrix,positioning,scopes,snakes,fadings}

\makeatletter
\tikzset{join/.code=\tikzset{after node path={%
\ifx\tikzchainprevious\pgfutil@empty\else(\tikzchainprevious)%
edge[every join]#1(\tikzchaincurrent)\fi}}}
\makeatother

\tikzset{>=stealth',every on chain/.append style={join},
         every join/.style={->}}
\tikzset{
    >=stealth',
    punkt/.style={
           rectangle,
           rounded corners,
           draw=black, very thick,
           text width=6.5em,
           minimum height=2em,
           text centered},
    pil/.style={
           ->,
           thick,
           shorten <=2pt,
           shorten >=2pt,}
}

\newcommand{\bea}{\begin{eqnarray}}
\newcommand{\eea}{\end{eqnarray}}
\newcommand{\be}{\begin{equation}}
\newcommand{\ee}{\end{equation}}
\newcommand{\nn}{\nonumber}
\newcommand{\Tr}{\textrm{Tr}}

\DeclareMathAlphabet{\mathpzc}{OT1}{pzc}{m}{it}

\usepackage{amsthm}
\usepackage{amsmath}
\usepackage{amsfonts}
\usepackage{amssymb}

\theoremstyle{definition}

\usetikzlibrary{arrows}
\newcommand{\midarrow}{\tikz \draw[>-stealth] (0,0) -- +(.1,0);}

\begin{document}
\hfill UUITP-27/24\\[25pt]

\begin{center} \Large
{\bf  Off-shell color-kinematics duality from codifferentials}
 \\[12mm] \normalsize
{\bf Maor Ben-Shahar${}^{a,b}$, Francesco Bonechi${}^{c}$ and Maxim Zabzine${}^{a,d}$} \\
[8mm]
{\small\it ${}^a$Department of Physics and Astronomy, Uppsala University,\\ Box 516, SE-75120 Uppsala, Sweden\\}
{\small\it ${}^b$ 
	Institut f\"ur Physik und IRIS Adlershof, Humboldt-Universit\"at zu Berlin,\\ 10099 Berlin, Germany\\
}
{\small\it ${}^c$ INFN Sezione di Firenze, Via G. Sansone 1,\\
 50019 Sesto Fiorentino, Firenze, Italy \\}
{\small\it ${}^d$Centre for Geometry and Physics, Uppsala University,\\
 Box 480, SE-75106 Uppsala, Sweden\\}
\end{center}
\vspace{7mm}

\begin{abstract}
We examine the color-kinematics duality within the BV formalism, highlighting its emergence as a feature of specific gauge-fixed actions. Our goal is to establish a general framework for studying the duality while investigating straightforward examples of off-shell color-kinematics duality. In this context, we revisit Chern-Simons theory as well as introduce new examples, including BF theory and 2D Yang-Mills theory, which are shown to exhibit the duality off-shell. We emphasize that the geometric structures responsible for flat-space color-kinematics duality appear for general curved spaces as well.
        \end{abstract}

\eject

\tableofcontents

\section{Introduction}

The color-kinematics duality \cite{Bern:2008qj,Bern:2010ue} reffers to an intriguing set of algebraic identities obeyed by the kinematic part of perturbative computations in a wide set of gauge theories.
A theory is said to obey the color-kinematics duality if there exists a decomposition of its scattering amplitudes in to sums of cubic diagrams, each associated with a propagator structure, a color factor, and a kinematic factor which obeys the same algebraic identities as the color factor.
The duality underlies the Bern-Carrasco-Johansson (BCJ) amplitude relations, and enables the amplitude building blocks from the gauge theory to be used in the double-copy construction, which produces amplitudes typically from theories of gravity. This double-copy construction generalizes the string-theory Kawai-Lewellen-Tye (KLT) relations \cite{Kawai:1985xq} in that it applies also to loops as well as a larger set of theories, see ref. \cite{Bern:2019prr} for a review.

Since the algebraic identities obeyed by the color factors stem from the Jacobi identity of the underlying gauge algebra, the color-kinematics duality suggests that there could be some kinematic Lie algebra underlying the analogous kinematic identities.  
In the self-dual sector of Yang-Mills theory such an algebra has been identified \cite{Monteiro:2011pc,Boels:2013bi}, and corresponds to an algebra of area preserving diffeomorphisms. In this case the numerators of the theory can be written with explicit commutators of generators of the diffeomorphisms, making the duality manifest. Similar constructions of differential operators have been identified for the non-linear sigma model \cite{Cheung:2016prv} and for Chern-Simons theory \cite{Ben-Shahar:2021zww}, where in the latter the kinematic algebra contains volume preserving diffeomorphisms. 
Further algebras have been identified in a non-abelian Navier-Stokes equation \cite{Cheung:2020djz} and two-dimensional integrable models \cite{Cheung:2022mix}.
In general, however, the kinematic algebra remains mysterious. 
Nonetheless, there exist direct methods for the construction of duality satisfying numerators \cite{Bjerrum-Bohr:2010pnr,Mafra:2011kj,Fu:2012uy,Mafra:2015vca,Bjerrum-Bohr:2016axv,Du:2017kpo,Chen:2017bug,Fu:2018hpu,Edison:2020ehu,He:2021lro,Bridges:2021ebs, Ahmadiniaz:2021fey,Cheung:2021zvb,Ahmadiniaz:2021ayd,Brandhuber:2021bsf,Cheung:2021zvb,Lee:2015upy,Bridges:2019siz,Chen:2022nei,Chen:2024gkj,Chen:2023ekh}, including at loop level also \cite{Bern:2010ue,Carrasco:2011mn,Bern:2012uf,Boels:2013bi,Bjerrum-Bohr:2013iza,Bern:2013yya,Nohle:2013bfa,Mogull:2015adi,Mafra:2015mja,He:2015wgf,Johansson:2017bfl,Hohenegger:2017kqy,Mafra:2017ioj,Faller:2018vdz,Kalin:2018thp,Duhr:2019ywc,Geyer:2019hnn,Edison:2020uzf,Casali:2020knc,DHoker:2020prr,Carrasco:2020ywq,Bridges:2021ebs,Edison:2023ulf} and for form factors \cite{Boels:2012ew,Yang:2016ear,Boels:2017ftb,Lin:2020dyj,Lin:2021qol,Lin:2021pne,Lin:2021lqo}. 

For pure Yang-Mills theory several attempts have been made at revealing the structure of the kinematic algebra by finding actions that directly produce duality-satisfying numerators \cite{Bern:2010yg,Tolotti:2013caa}, with recent work \cite{Ben-Shahar:2022ixa} identifying a deformed Lagrangian for the NMHV sector of Yang-Mills theory (in which combinatorial algebras for the numerators exist also \cite{Chen:2019ywi,Chen:2021chy}). These constructions rely on fixing a particular gauge, and searching for contact terms that correct the Jacobi identities of the numerators. The existence of a kinematic Lie algebra indeed appears to be a gauge-dependent statement \cite{Bonezzi:2023pox,Armstrong-Williams:2024icu}, thus it is not surprising that a particular gauge choice should be needed. In a related setting, there has been much recent work on developing the homotopy-algebraic interpretation of the color-kinematics duality and double copy \cite{Reiterer:2019dys, Borsten:2021hua,Borsten:2021gyl,Borsten:2022ouu,Borsten:2022vtg,Borsten:2023reb,Borsten:2023ned,Bonezzi:2022yuh,Bonezzi:2022bse,Bonezzi:2023lkx,Bonezzi:2023pox,Bonezzi:2023ciu,Bonezzi:2024dlv,Escudero:2022zdz,Szabo:2023cmv}. In this paper we study off-shell color-kinematics duality using the BV formalism, which is closely related to the aforementioned topics. 

  Our approach, inspired by refs. \cite{Ben-Shahar:2021zww} and \cite{Reiterer:2019dys}, focuses on investigating a specific gauge fixing within the BV formalism.  In ref. \cite{Ben-Shahar:2021zww} the off-shell color-kinematics duality for the Chern-Simon theory has been discussed and it has been observed that the  Lorenz gauge is imposed by means of a second-order operator $d^\dagger$ which generates the Lie bracket for the kinematic algebra. 
Thus, the off-shell color-kinematics duality for Chern-Simons theory is the direct consequence of the second-order property of the gauge-fixing operator.
We develop and 
  formalize this observation to a general framework within the BV formalism.  
  Let us outline our idea schematically, with the details provided in the main text.
   Within the BV formalism many gauge theories can be recast as formal Chern-Simons theories defined on the maps ${\cal C}~\rightarrow~\mathfrak{g}[1]$, with action 
   \bea\label{BV-general-CS-intro}
          S_{f-CS} = \int d\mu ~ \Big (\frac{1}{2} {\cal A}^a D {\cal A}^b \delta_{ab} + \frac{1}{6} {\cal A}^a {\cal A}^b {\cal A}^c f_{abc} \Big )~,
        \eea 
  where $\mathfrak g$ is a compact Lie algebra, with totally antisymmetric structure constants $f_{abc}$, and $({\cal C}, D, \int d\mu )$ is a graded commutative differential algebra together with an integral $\int d\mu$ of degree $-3$ (modelling  on the example
   of differential forms ${\cal C} = \Omega^\bullet (\Sigma_3)$ with $D=d$ being de Rham differential and $\int d\mu$ is standard integration 
    of forms). In  general the structure of $({\cal C}, D, \int d\mu )$ may be quite involved and  ${\cal A}^a$ are not necessarily freely generated superfields. The gauge fixing  in the BV formalism is defined by specifying a Lagrangian submanifold in the space of fields with respect 
     to the odd symplectic structure
     \be
      \omega_{f-CS} = \int d\mu ~ \delta {\cal A}^a \wedge \delta {\cal A}^b  \delta_{ab} \ ,
     \ee
   and it can 
     be done by choosing a suitable operator $D^\dagger$ of degree $-1$ such that it
       satisfies $(D^\dagger)^2=0$ and (\ref{basic-prop-3}), a version of Stokes' theorem with respect to $\int d\mu$. In the examples we will consider, $\{ D , D^\dagger \}=\square$ is a 
        nice elliptic operator, e.g. the Laplace operator.
        The  Lagrangian submanifold is defined by the condition ${\cal D}^\dagger {\cal A}^a=0$
      which is a sort of generalized Lorenz gauge. If we evaluate the action
       (\ref{BV-general-CS-intro}) on ${\rm Im} (D^\dagger)$ (for the moment we ignore the difference  between $\ker$ and ${\rm Im}$) then the 
        gauge-fixed action can be written as 
  \bea\label{gaged-fixed-intro}
     S_{CS} = \frac{1}{2} \langle {\cal A}^a, \square {\cal A}^b \rangle \delta_{ab} +  \frac{1}{6}
     \langle {\cal A}^a, \{ {\cal A}^b , {\cal A}^c\} \rangle  f_{abc}~,
    \eea
    where on ${\rm Im} (D^\dagger)$ we defined the bracket $\{ {\cal A}^b , {\cal A}^c\} = D^\dagger ({\cal A}^b , {\cal A}^c)$ and the invariant pairing $\langle~, ~\rangle$, see eqn. \eqref{pairing-Cs-1f}.
     If $D^\dagger$ is a second-order operator then the bracket $\{~,~\}$ is a Lie bracket. 
 If we restrict ourselves to flat space  $\mathbb{R}^d$ and choose $\square$ to be the Laplace operator acting diagonally on forms then  in momentum 
  representation $\square$ corresponds  to multiplication by an overall  $p^2$-factor. Thus we obtain off-shell color-kinematics duality for such a theory. 
   If the order of $D^\dagger$ is higher than two then the bracket $\{~,~\}$ is a two bracket for some $L_\infty$ algebra. These structures are not associated with any obvious symmetry of the gauge-fixed action (\ref{gaged-fixed-intro}).
   It is important to emphasize 
that the emergence of the kinematic Lie algebra from the gauge-fixing data is not limited to the flat case of $\mathbb{R}^d$, although its manifestations in the compact case will not be addressed in detail in the present paper.

The paper is organized as follows: In Section \ref{s:heuristic} we review tree-level color-kinematics duality for Chern-Simons theory in the Lorenz gauge.  This includes a rephrasing of the results from \cite{Ben-Shahar:2021zww}, stressing their algebraic meaning.
  In Section \ref{s:proper} we reformulate these results independently of the momentum representation, incorporating also the full spectrum of ghosts of Chern-Simons theory. The color-kinematics duality relies 
    on the existence of two structures: a Lie bracket and an invariant metric defined on the gauge-fixed fields. In this section we also include a discussion of Chern-Simons theory on compact manifolds. 
 Section \ref{s:formal} contains generalizations of the two previous sections, fleshing out conditions under which
  a formal Chern-Simons theory will have  color-kinematics duality off-shell. 
 In  Sections \ref{s:BF} and \ref{s:YM} we give two explicit examples of theories which can be written as formal Chern-Simons 
  theories and which exhibit color-kinematics duality off-shell.  In Sections \ref{s:BF} we consider 4-dimensional BF-theory, we present 
   both formal discussions and concrete calculations for tree numerators in flat space. In  Section \ref{s:YM} we show that 2-dimensional 
    Yang-Mills theory has off-shell color-kinematics duality in a specific gauge. 
 Section \ref{s:summary} presents  the summary of these results and their possible extensions. 
  In Appendix \ref{app-A} we collect notations and conventions for differential forms in $\mathbb{R}^d$ and in Appendix \ref{app-B} we
   summarize the basic algebraic definitions of  the order of operators and the definitions of derived brackets.  
 
\section{Tree diagrams in Chern-Simons theory}\label{s:heuristic}

In this section we consider tree-level diagrams in Chern-Simon theory. We work with off-shell truncated correlation functions, because putting the external states on-shell will trivialise the amplitudes. However, off-shell diagrams still serve as a nice toy model for the general picture which we will discuss later on. We closely follow ref. \cite{Ben-Shahar:2021zww}. 
 
 We consider Chern-Simons theory defined over $\mathbb{R}^3$ with the standard metric $g_{\mu\nu} = {\rm diag}(1,1,1)$ and a compact Lie algebra $\mathfrak{g}$.
   We define the classical Chern-Simons action as follows  
 \be\label{CS-classical}
 S_{CS} = \int\limits_{\mathbb{R}^3} \Tr \Big ( A dA + \frac{2}{3} A^3 \Big ) = \int\limits_{\mathbb{R}^3} \Big ( \frac{1}{2} A^a dA^b \delta_{ab} + \frac{1}{6} A^a A^b A^c f_{abc} \Big )~,
\ee
where $A$ is a Lie algebra valued one-form on $\mathbb{R}^3$, and multiplication of forms is assumed to be the wedge product. Here  $\delta_{ab}$ and $f_{abc}$ are the invariant metric and structure constants for the Lie algebra $\mathfrak{g}$,
  \be
  {\rm Tr}(T_a T_b)= \frac{1}{2} \delta_{ab}~,~~~~~[T_a, T_b] = f_{ab}^{~~c} T_c~.
  \ee
   In order to calculate tree diagrams we impose the Lorenz gauge-fixing condition $\partial^\mu A_\mu=0$
   on $A$, or, in a coordinate free version, $d^\dagger A=0$, where $d^\dagger$ is the codifferential defined by the metric.  In this gauge $A$ can be written as follows
\be\label{solved-gauge}
 A^a_\mu (x) = \partial^\nu \Lambda^a_{\nu\mu} (x) = \epsilon_{\mu}{}^{\nu\rho} \partial_\nu \xi^a_\rho (x) = 
\frac{1}{(2\pi)^{3/2}} \int d^3p~ e^{i x\cdot p} \epsilon_{\mu}{}^{\nu\rho}  p_\nu \xi^a_\rho (p) ~,
\ee
 where we would  assume $p^\mu \xi_\mu (p)=0$ in order to avoid additional ambiguities and $\epsilon^{\mu\nu\rho}$ is the totally antisymmetric Levi-Civita symbol with $\epsilon^{123}=1$. Let us evaluate the Chern-Simons 
  action (\ref{CS-classical}) on $A$ expressed as above. Using the identity
 \bea
  \epsilon_{\mu\nu\rho} \epsilon^{\lambda\sigma\rho} = \delta^\lambda_\mu \delta^\sigma_\nu - \delta^\sigma_\mu \delta^\lambda_\nu \ , 
 \eea
the kinetic term has the following form
 \begin{align}
\int\limits_{\mathbb{R}^3}  A^a d A^b \delta_{ab} 
&= \frac{1}{3!} \int\limits_{\mathbb{R}^3}  d^3x ~\epsilon^{\mu\nu\rho} A^a_\mu (x) \partial_\nu A^b_\rho (x) \delta_{ab}  \nonumber \\
&= \frac{1}{3!} \int\limits_{\mathbb{R}^3}  d^3p~\xi^a_\mu(p) \xi^b_\nu(-p) \Delta^{\mu\nu}(p) p^2 \delta_{ab} \ 
\end{align}
in momentum space, where we introduced the following operator (with notation following ref. \cite{Ben-Shahar:2021zww})
\bea
 \Delta^{\mu\nu}(p) = - i \epsilon^{\mu\nu\rho} p_\rho~.
\eea
The interaction term can be expressed as follows 
\begin{align}
 \int A^a  A^b A^c  f_{abc} 
 &= \frac{1}{3!} \int d^3x~ \epsilon^{\mu\nu\rho} A^a_{\mu} (x) A^b_{\nu} (x) A^b_{\rho}(x)  f_{abc} \nonumber  \\
&= \frac{1}{3!} \int d^3p_1~ d^3p_2~ d^3p_3~ \Big(\xi^a_{\sigma} (p_1) \xi^b_{\gamma}(p_2) \xi^c_{3\phi}(p_3) \nonumber\\
&\hspace{2.5cm} \times F^{\sigma\gamma}_{~~~\rho} (p_1, p_2; -p_3) \Delta^{\rho\phi}( -p_3) f_{abc}\Big)~,
\end{align}
 where we introduced the kinematic structure constants 
 \bea\label{CS-inv-metric-prop}
 i F^{\mu\nu}_{~~~\rho} (p_1, p_2;  -p_3) \Delta^{\rho\gamma}(p_3) = 
  \frac{i}{(2\pi)^{3/2}}
   \epsilon^{\mu\sigma\lambda} p_{1\lambda} \epsilon^{\nu\delta\gamma} p_{2\gamma} \epsilon^{\gamma \rho \phi} p_{3\phi} \epsilon_{\sigma\delta\rho} \delta (p_1\!+\!p_2\!+\!p_3),
\eea
which are totally antisymmetric with respect to permutations of $(\mu, p_1)$, $(\nu, p_2)$, $(\gamma, p_3)$.
  
We can derive momentum-space Feynman rules in Lorenz gauge from the Chern-Simons action with the $\xi^\mu$ fields. 
  The propagator is defined by the following expression 
\be\label{CS-prop-pic}
\begin{tikzpicture}[baseline={(0, 0cm)}]
\draw[thick] (-1,0) -- node {\midarrow} (1,0);
\node [font=\tiny] at (-1.0,-0.25) {($a$, $\mu$)};
\node [font=\small] at (0.0,0.25) {$p$};
\node [font=\tiny] at (1.0,-0.25) {($b$, $\nu$)};
\end{tikzpicture} 
=\ 
      \frac{\Delta_{\mu\nu}(p)}{p^2} \delta^{ab}~,
  \ee
   where we used the identity 
 \bea
  \Delta^{\mu\nu} (p) \Delta_{\nu\sigma} (p) \xi^\sigma(p)=  p^2 \xi^\mu(p)
 \eea
  which holds because the fields satisfy  $p^\mu \xi_\mu(p)=0$.
 The trivalent vertex is defined by
\bea\label{CS-trivalent-pic}
\begin{tikzpicture}[baseline={(0, 0cm)}]
\begin{scope}[very thick, every node/.style={sloped,allow upside down}]
---\draw[thick] (-1.5,1) -- node {\midarrow} (0,0);
\draw[thick] (-1.5,-1) -- node {\midarrow} (0,0);
\draw[thick] (0,0) -- node {\midarrow} (1.5,0);
\node [font=\tiny] at (-1.5,1.25) {($a$, $\mu$)};
\node [font=\small] at (-0.75,0.85) {$p_1$};
\node [font=\small] at (-0.75,-0.85) {$p_2$};
\node [font=\small] at (0.75,0.25) {$p_3$};
\node [font=\tiny] at (-1.5,-1.25) {($b$, $\nu$)};
\node [font=\tiny] at (1.5,0.25) {($c$, $\gamma$)};
\end{scope}
\end{tikzpicture} 
=
- i F^{\mu\nu}_{~~~\rho} (p_1, p_2;  -p_3) \Delta^{\rho\gamma}(-p_3) f_{abc}
\ ,
 \eea
 where the arrows indicate the direction of the flow of momentum.
These Feynman rules apply to the vector field and are sufficient for tree diagrams, while loops will additionally require the inclusion of ghosts. These will be included in the next section. 
 
  Next we want to understand the algebraic meaning of the Feynman rules in this gauge. 
 Let us consider the collection of  vector fields
 \bea
 L^\mu (p) = \frac{1}{(2\pi)^{3/2}} e^{ip\cdot x} \epsilon^{\mu\nu\rho} p_\nu \frac{\partial}{\partial x^\rho}~,
\eea
 which we can regard as a basis of the Lie algebra of divergenceless vector fields 
\bea
 V^\rho (x)  \frac{\partial}{\partial x^\rho}  = \frac{1}{(2\pi)^{3/2}} \int d^3p~ e^{ip\cdot x} \epsilon^{\mu\nu\rho} p_\nu \xi_\mu (p)  \frac{\partial}{\partial x^\rho}
  =  \int d^3p~ \xi_\mu(p) L^\mu(p)~.
\eea
 Here again to avoid the redundancy of description we assume that $p^\mu \xi_\mu (p)=0$.  
  They form a Lie algebra with the structure constants
\bea
 [L^\mu (p_1), L^\nu(p_2) ] = F^{\mu\nu}_{~~~\rho} (p_1, p_2; p_3) L^\rho (p_3)~,
\eea
where
\bea
 F^{\mu\nu}_{~~~\rho} (p_1, p_2; p_3) = \frac{i}{(2\pi)^{3/2}} \epsilon^{\mu\sigma\lambda} p_{1\lambda} \epsilon^{\nu\delta\gamma} p_{2\gamma} \epsilon_{\sigma\delta\rho} \delta (p_1+p_2-p_3)~ 
\eea
are the kinematics structure constants defined in (\ref{CS-inv-metric-prop}) which also appear as the kinematics part of the trivalent vertex (\ref{CS-trivalent-pic}). 
  These structure constants satisfy  the Jacobi identity
\bea
 F^{\nu\rho}_{~~~\sigma} (p_2, p_3; p_4) F^{\sigma\mu}_{~~~\gamma} (p_4, p_1; p_5)  +
 F^{\rho\mu}_{~~~\sigma} (p_3, p_1; p_4) F^{\sigma\nu}_{~~~\gamma} (p_4, p_2; p_5) \nonumber \\
 +
 F^{\mu\nu}_{~~~\sigma} (p_1, p_2; p_4)  F^{\sigma\nu}_{~~~\gamma} (p_4, p_3; p_5) =0~. 
\eea
    The kinematic part  $\Delta^{\mu\nu}(p)$ of the numerator of the propagator (\ref{CS-prop-pic}) defines a pairing that is invariant  
     due to antisymmetry properties of (\ref{CS-inv-metric-prop}). 
    Thus if we look at any tree diagram and we strip off the $p^2$-factors in the denominators, we are left with the color part made of the contraction of structure constants $f$ and metric $\Tr$ for Lie algebra $\mathfrak{g}$, as well as the kinematic part, made of contractions of the structure constants $F$ and pairing $\Delta$ for the Lie algebra of divergenceless vector fields. 
       Both factors satisfy the Jacobi identities and this is exactly the color-kinematics duality.

\section{Chern-Simons theory beyond trees}\label{s:proper}
 
In this section we explain
 the geometrical origin of the kinematic Lie algebra  that appeared in the previous Section. We stress that the construction does not require that we work in ${\mathbb R}^3$, in particular 
 it is independent of the momentum representation.
  Moreover, we include ghosts so that the arguments presented here generalise to all diagrams including loops.
 
  First we recall a few properties of Lie brackets of vector fields on smooth manifolds. Consider a smooth manifold $\Sigma_d$ with metric  $g_{\mu\nu}$,  then the standard Lie bracket of vectors fields can be rewritten as follows
 \bea
 &&  \{ V, W\}^\mu = V^\rho \partial_\rho W^\mu - W^\rho \partial_\rho V^\mu = V^\rho \nabla_\rho W^\mu - W^\rho \nabla_\rho V^\mu  \nn \\
   &&=
    \nabla_\rho (V^\rho W^\mu - W^\rho V^\mu) - (\nabla_\rho V^\rho) W^\mu + (\nabla_\rho W^\rho) V^\mu~,
 \eea
  where $\nabla_\rho$ is the Levi-Civita covariant derivative for the metric $g$. If we lower all indices we obtain the Lie bracket on one-forms
   \bea
    \{ V, W\}_\mu = 
    \nabla^\rho (V_\rho W_\mu - W_\rho V_\mu) - (\nabla^\rho V_\rho) W_\mu + (\nabla^\rho W_\rho) V_\mu~.
   \eea
   Remembering the definition of the codifferential $d^\dagger$ as contraction with $\nabla^\rho$  we can rewrite the above Lie bracket on one-forms as follows
    \bea \label{eqn:oneformsbracket}
     \{ V, W \} = d^\dagger (V \wedge W)  - (d^\dagger V) \wedge W  + V\wedge  (d^\dagger W) ~.
    \eea
Thus $d^\dagger$ defines a Lie algebra of one forms that is isomorphic through the metric to the algebra of vector fields. The fundamental fact here is that the bracket (\ref{eqn:oneformsbracket}) satisfies the Jacobi identity since $d^\dagger$ is a second-order linear  operator on the commutative graded algebra of forms $(\Omega(\Sigma_d),\wedge)$. 
 
 The subspace $\ker (d^\dagger)\subset\Omega^1(\Sigma_d)$ is a subalgebra that is isomorphic to the Lie algebra of divergenceless vector fields. Since $d^\dagger{}^2=0$, the image of the codifferential ${\rm Im}(d^\dagger)$ is a Lie subalgebra of $\ker (d^\dagger)$ that corresponds to the Lie subalgebra of vector fields of the form $V^\mu \partial_\mu=
          \nabla_\rho \Lambda^{\rho\mu}\partial_\mu$ for some bi-vector field $\Lambda = \Lambda^{\rho\mu} \partial_\rho \wedge \partial_\mu$. 
   In flat space $\mathbb{R}^d$, $\ker d^\dagger$ and ${\rm Im}d^\dagger$ coincide: the kinematic Lie algebra that we considered in the previous Section for ${\mathbb R}^3$ is then ${\rm Im}(d^\dagger)$ endowed with the bracket (\ref{eqn:oneformsbracket}) and makes sense for Chern-Simons theory defined on any smooth $\Sigma_3$. 

Let us now consider the pairing. Let $\omega_1,\nu_1\in{\rm Im} (d^\dagger)\subset\Omega^1(\Sigma_3)$, then
 \bea\label{pairing-Cs-1f}
  \langle \omega_1, \nu_1 \rangle = \int \omega_1 \wedge \xi_2~,
 \eea
 where $\nu_1=d^\dagger\xi_2$. 
  This pairing is invariant under shifts of $\xi$ by any form which is $d^\dagger$-closed, 
   because $d^\dagger$ satisfies
   \begin{align}\label{stokes_codiff}
   \int d^\dagger \omega_p~  \alpha_q = (-1)^p\int \omega_p ~d^\dagger \alpha_q~.
   \end{align}
Moreover, this pairing is non-degenerate, symmetric, and invariant with respect to the bracket (\ref{eqn:oneformsbracket}). Indeed, let $\omega_1,\nu_1,\alpha_1\in{\rm Im}(d^\dagger)$; the bracket reads \bea\label{eqsimplebracketCS}
\{ \omega_1, \alpha_1 \} = d^{\dagger} (\omega_1 \wedge \alpha_1)~
\eea
so that we can write 
\be
\int \omega_1 \wedge \alpha_1 \wedge \nu_1 =  \langle \omega_1 , \{\alpha_1 , \nu_1\} \rangle = \langle \nu_1 , \{\omega_1 , \alpha_1\} \rangle~.
\ee
Therefore the evaluation of the Chern-Simons action (\ref{CS-classical}) on ${\rm Im} (d^\dagger)$ leads to the following form 
   \be
   S_{CS} = \frac{1}{2}  \langle A^a, \square A^b \rangle \delta_{ab}  + \frac{1}{6}  \langle A^a, \{ A^b, A^c \} \rangle f_{abc} ~,
   \ee
  where $\square = d d^\dagger + d^\dagger d$ is the Laplace operator. This formulation of the gauge-fixed action is valid for general $\Sigma_3$ and is
  equivalent to the momentum-space action studied in the previous section on $\mathbb{R}^3$.
  
This discussion can be extended to the full theory including ghosts, that are needed to deal with loop diagrams. For this we need to enlarge the space of fields, and we go through the BV treatment of  Chern-Simons theory to do this consistently.  
  Following the AKSZ construction \cite{Alexandrov:1995kv} for Chern-Simons theory we consider the maps
  \be
   T[1] \Sigma_3 ~\longrightarrow~\mathfrak{g}[1]~,
  \ee
   which correspond to  the collection of Lie-algebra valued differential forms 
 \be
  {\cal A}^a = A^a_0 + A^a_1 + A^a_2 + A^a_3~,
 \ee  
where the subscript specifies the degree of the form.

We can understand ${\cal A}^a (x, \theta)$ as a superfield whose expansion in $\theta\equiv dx$ can be identified with differential forms of different degrees. In terms of this superfield the BV action for the Chern-Simons theory is given by
    \be\label{CS-BV-1}
    S_{CS}= \int d^3 x d^3 \theta ~ \Big ( \frac{1}{2} {\cal A}^a d {\cal A}^b \delta_{ab} + \frac{1}{6} {\cal A}^a {\cal A}^b {\cal A}^c  f_{abc} \Big )~, 
    \ee
  where $d = \theta^\mu \frac{\partial}{\partial x^\mu}$ is the de Rham differential realized on superfields.   
 This action satisfies the classical master equation with respect to the following odd symplectic form
 \be\label{sympform-CS}
  \omega_{CS} = \int d^3 x d^3 \theta ~ \delta  {\cal A}^a \wedge  \delta {\cal A}^b \delta_{ab}~,
 \ee
 where $\delta\mathcal{A}^a$ are functional differentials.
  The gauge fixing corresponds to a choice of Lagrangian submanifold with respect to this odd symplectic structure, namely a restriction of ${\cal A}^a$ to the largest subset of fields such that the symplectic form vanishes.
   If we choose ${\cal A}^a$ to be in ${\rm Im} (d^\dagger)$ then the symplectic form (\ref{sympform-CS}) is zero (modulo zero modes 
    which we discuss later).  This gauge fixing can be implemented either by introducing Lagrange multipliers and anti-ghosts together with their anti-fields, or it can be implemented directly by requiring all fields to be in  ${\rm Im} (d^\dagger)$. 
These two treatments are equivalent and for the sake of clarity we just assume that all fields are in ${\rm Im} (d^\dagger)$ and avoid the introduction of any additional fields.
Let us evaluate the BV action (\ref{CS-BV-1}) on ${\rm Im} (d^\dagger)$.  The Lie algebra on one-forms can be extended to the following graded Lie algebra structure on $\Omega^\bullet(\Sigma_3)$ (whose degree is shifted by $1$)
 \be\label{def-bracket-forms}
   \{ \omega_p, \alpha_q \} = d^\dagger (\omega_p \alpha_q)  - (d^\dagger \omega_p) \alpha_q  -  (-1)^p \omega_p (d^\dagger \alpha_q)~.
 \ee
Using the metric it can be mapped to the Schouten-Nijenhuis bracket on multivector fields. As before, on ${\rm Im}(d^\dagger)\subset\Omega^\bullet(\Sigma_3)$ the Lie subalgebra admits the symmetric, non degenerate invariant pairing 
\be\label{pairing-Cs}
      \langle \omega_p, \alpha_q \rangle = \int \omega_p \wedge \xi_{q+1}~,
\ee
where $\alpha_q = d^\dagger \xi_{q+1}$.  To summarize: the differential forms in  ${\rm Im}(d^\dagger)$ (with the shifted degree) give rise to
 the graded version of quadratic Lie algebra, with the Lie bracket $\{~,~\}$ and the invariant pairing $\langle~,~\rangle$ defined in \eqref{def-bracket-forms} and \eqref{pairing-Cs} respectively.
 
 Combining all this together we can rewrite the Chern-Simons BV action
   evaluated on ${\rm Im}(d^\dagger)$ as follows
 \be
 S_{CS}=  \frac{1}{2}\langle  {\cal A}^a,  \square {\cal A}^b \rangle  \delta_{ab} + \frac{1}{6}\langle  {\cal A}^a, \{ {\cal A}^b , {\cal A}^c \} \rangle  f_{abc}  ~,
 \ee
  where $\square$ is the Laplace operator.
  Since $\square$ is invertible on ${\rm Im}(d^\dagger)$ we can define 
    Feynman rules using all this algebraic data very much in analogy with the tree diagrams studied earlier. The appearance of the kinematic Lie bracket in the action makes color-kinematics duality manifest. However, we have yet to deal with the zero modes that may appear in intermediate diagrams, and explain how they affect the color-kinematics duality. This will be discussed in the next subsection.

 \subsection{Chern-Simons theory on a compact space}\label{s:CS-compact}
The fundamental property of the Lorenz gauge fixing is expressed by the fact that the de Rham differential $d$, defining the kinetic term, is invertible when restricted to gauge-fixed configurations. We have to properly take care of the harmonic forms, the zero modes, that otherwise would spoil this property. In $\mathbb{R}^3$ using Fourier transform we can choose a basis for forms in  $\textrm{Im}(d^\dagger)$ and work out structure constants explicitly, see Section \ref{s:heuristic}. Zero modes appear then as poles in the Feynman propagators in the momentum representation, and correspond to regions of kinematic space where intermediate particles go on-shell. On compact manifolds we do not have such a simple tool: the zero modes must be treated as external fields that are not integrated. In the compact case this is a standard procedure based on the Hodge decomposition (see for instance \cite{Cattaneo_2009}). We show here that this decomposition fits nicely into the general algebraic structures identified previously.
    
We first review some important algebraic properties of differential forms.
  On a compact manifold $\Sigma_d$ equipped with a metric we have the Hodge decomposition of the differential forms into 
   three orthogonal spaces
   \be
    \Omega^\bullet = \ker (\square) \oplus {\rm Im} (d) \oplus {\rm Im}(d^\dagger) \ ,
   \ee
    and thus we have 
    \be
      \ker(d^\dagger) = \ker (\square) \oplus  {\rm Im}(d^\dagger)~.
    \ee
Let us denote elements of $ \ker(d^\dagger)$ as $a+{\cal A}$, with $a$ corresponding to the harmonic part and ${\cal A}$ is in  ${\rm Im}(d^\dagger)$.
 The restriction of the bracket (\ref{def-bracket-forms}) to $\ker(d^\dagger)$ 
 \be\label{bracket-ker}
  \{ a^a + {\cal A}^a, a^b + {\cal A}^b \} = d^\dagger\Big  ( (a^a + {\cal A}^a )(a^b + {\cal A}^b) \Big )
 \ee
 still forms a Lie algebra and ${\rm Im}(d^\dagger)$ is a subalgebra. Thus we observe
 \be
  \{  \ker(d^\dagger),  \ker(d^\dagger) \} \subset {\rm Im}(d^\dagger)~.
 \ee
 Next, let us define the pairing on $\ker(d^\dagger)$. On harmonic elements we define it to be zero
 \be\label{pairing_zero_modes}
  \langle a^a, a^b \rangle =0~.
 \ee
 On other elements we define it as follows
\bea
&& \langle a^a , {\cal A}^b \rangle = \langle {\cal A}^b, a^a \rangle = \int a^a \wedge \xi^b~,\\
&&  \langle {\cal A}^a , {\cal A}^b \rangle =  \int {\cal A}^a \wedge \xi^b~,
\eea
 where ${\cal A}^b = d^\dagger \xi^b$. The pairing is well-defined in the sense that it does not depend on the choice of $\xi$.
  This pairing is symmetric and degenerate due to (\ref{pairing_zero_modes}).  The crucial property of this pairing is that it is invariant with respect to the bracket (\ref{bracket-ker})
    \be
      \langle \{  a^a + {\cal A}^a , a^b + {\cal A}^b \},  a^c + {\cal A}^c \rangle = - \langle a^b + {\cal A}^b, \{a^a+{\cal A}^a, a^c
       + {\cal A}^c \} \rangle~. 
     \ee
   These properties are true for any compact manifold equipped with a metric. 
   
   Now let us go back to Chern-Simons theory on $\Sigma_3$.  Let us evaluate the Chern-Simons action (\ref{CS-BV-1}) on 
    $\ker(d^\dagger)$ and we use the same notations as before, namely $a$ is harmonic and ${\cal A}$ is in ${\rm Im}(d^\dagger)$. 
    Using the above definitions of the bracket and the pairing we get the following
\bea
 && S_{CS}=  \frac{1}{2}\langle  {\cal A}^a,  \square {\cal A}^b \rangle  \delta_{ab}  +
 \frac{1}{6}\langle  {\cal A}^a, \{ {\cal A}^b , {\cal A}^c \} \rangle  f_{abc}  \\ 
 &&  + \frac{1}{2} \langle  a^a, \{ {\cal A}^b , {\cal A}^c \} \rangle  f_{abc} +  \frac{1}{2} \langle  a^a, \{ a^b , {\cal A}^c \} \rangle  f_{abc} +\frac{1}{6}\langle  a^a, \{ a^b , a^c \} \rangle  f_{abc} ~. \nn
\eea
  The harmonic fields $a$ act as external sources in perturbative calculations of the partition function. In addition, all Feynman diagrams, including those with vertices with 
    harmonic external states, have a clear algebraic meaning in terms of the Lie algebra on $\ker(d^\dagger)$. Importantly, the harmonic forms do not appear on internal legs, as the subspaces  $\ker (\square)$ and ${\rm Im}(d^\dagger)$ are orthogonal. In the more general theories that we are going to discuss in the next section the splitting between zero modes and the gauge-fixed subspace is not so simple.

\section{Formal Chern-Simons theory}\label{s:formal} 

In this section we abstract the properties that were important for color-kinematics duality in Chern-Simons theory, laying the groundwork for subsequent sections where we show that color-kinematics duality for BF theory and 2D Yang-Mills can be understood in this language.  
 We follow the AKSZ construction \cite{Alexandrov:1995kv} of formal Chern-Simons theory, 
  for different versions of the formalism one can consult  \cite{Cattaneo_2009} and \cite{Bonechi:2010tbl}. 
  
  Consider the differential graded algebra, or dga for short, $({\cal C}, D)$ with graded commutative multiplication and  differential  $D: {\cal C}^\bullet \rightarrow {\cal C}^{\bullet+1}$, together with integration  $\int d\mu: \cal C \rightarrow \mathbb{R}$ as a non-degenerate pairing of degree  $-3$, meaning $\int d\mu\ a \neq 0$ implies $|a|=3$. The integration obeys
   \bea
 \int d\mu \ Da = 0~,
   \eea
   or equivalently 
   \bea\label{D-int-bypart}
    \int d\mu ~Da b = -(-1)^{|a|} \int d\mu~a Db~.
   \eea
Altogether, the dga with the pairing $\int d\mu$ forms a differential Frobenius algebra. In Chern-Simons, as we saw above, $\mathcal{C}$ was given by the differential forms $\Omega^\bullet (\Sigma_3)$ with the standard wedge product, $D$ the de Rham differential $d$, and the pairing $\int d\mu$ is integration of differential forms.
We assume that $\cal C$ is bounded, {\it i.e.} it has the following structure,
\begin{equation}\label{formal-CS-degree}
\begin{tikzcd}
 & {\cal C}^{-k} \arrow[r,"D"] & {\cal C}^{-k+1}  \arrow[r, "D"]  &    ....   \arrow[r, "D"]       & {\cal C}^{k+2}     \arrow[r, "D"]   & {\cal C}^{k+3}
\end{tikzcd}
\end{equation}
for some positive integer $k$.
The action for the formal Chern-Simons theory has the following form
        \bea\label{BV-general-CS}
          S_{CS} = \int d\mu ~ \Big (\frac{1}{2} {\cal A}^a D {\cal A}^b \delta_{ab} + \frac{1}{6} {\cal A}^a {\cal A}^b {\cal A}^c f_{abc} \Big )~,
        \eea 
     where ${\cal A}^a$ is a Lie-algebra valued element of $\mathcal{C}$ and has overall degree $1$.
The corresponding BV symplectic form is given by 
      \bea
        \omega_{CS} = \int d\mu ~ \delta {\cal A}^a \wedge \delta {\cal A}^b \delta_{ab}~.
      \eea
       The above action satisfies the classical master equation by construction. 
      For example,  4D Yang-Mills theory can be recast in this form, see ref. \cite{MR2778558}.

  Next, let us discuss the gauge fixing for formal Chern-Simons theory. We would like to formalize the features which we have 
   observed in the actual Chern-Simons theory in previous sections. We therefore want to introduce the operator $D^\dagger$ of degree $-1$ which 
    satisfies the following properties:
    \bea
    &{\rm Property~1}~~~~~~~ & (D^\dagger)^2=0~,\label{basic-prop-1}\\
     & {\rm Property~2}~~~~~~~  &  \{ D, D^\dagger\} = \square~, \label{basic-prop-2}\\
     & {\rm Property~3}~~~~~~~   & \int d\mu ~D^\dagger (a) b =   (-1)^{|a|} \int d\mu~ a D^\dagger (b)~,\label{basic-prop-3}
    \eea
     where $\square$ is some reasonable elliptic operator 
     and here $a$, $b$ are any elements of ${\cal C}$ of definite degree.    
     For the moment we can consider (\ref{basic-prop-2}) as a definition of $\square$.
     Note that  $\square$ is hermitian with respect to the pairing defined by $\int d\mu$ using  (\ref{D-int-bypart}) and the choice of signs in (\ref{basic-prop-3}). 
 
   Let us assume that we can define a projection ${\rm P}_{\ker \square}$ on $\ker(\square)$ which is a linear subspace of ${\cal C}$.     
        If we take  an arbitrary element $a$ 
      we can rewrite property 2 as follows
       \bea
         D D^\dagger  (a  - {\rm P}_{\ker \square} a)+ D^\dagger D (a - {\rm P}_{\ker \square} a)= \square (a - {\rm P}_{\ker \square} a)~.
       \eea
      We can invert the $\square$ operator on the complement of its kernel
        and therefore
        \bea\label{decomposition}
         a = {\rm P}_{\ker \square} a + D \frac{D^\dagger}{\square} (a  - {\rm P}_{\ker \square} a) +  D^\dagger   \frac{1}{\square}   D (a  - {\rm P}_{\ker \square} a)~,
        \eea
      where we use the fact that   $\square$ commutes with both $D$ and $D^\dagger$ operators and  so 
       the operators $D$, $D^\dagger$ preserve $\ker (\square)$ and its complement. If we denote with 
       \begin{equation}
       \label{propagator}
       K = \frac{1}{\square}(1-P_{\ker\square})D^\dagger
       \end{equation}
we can write (\ref{decomposition}) as
\begin{equation}\label{homotopy}{\rm id} - P_{\ker\square} = D K + KD\;.
\end{equation}       
Since $[\square,D]=0$, $(\ker\square,D)$ is a subcomplex of the full $({\cal C},D)$ and $K$ defines a cochain homotopy between the two complexes. In particular the subcomplex $(\ker\square,D)$ computes the same cohomology of $({\cal C},D)$. We will refer to $\ker\square$ as the space of zero modes. 
       We thus have the following decomposition of the space of fields
    \bea\label{general_decomposition}
     {\rm ker} (\square) \oplus (1-{\rm P}_{\ker \square}){\rm Im} (D) \oplus  (1-{\rm P}_{\ker \square}){\rm Im}(D^\dagger)~.
    \eea  
      With respect to the pairing $\int d\mu$ we have the following orthogonality property
      \bea
    &&   ( {\rm ker} (\square))^\perp =(1-{\rm P}_{\ker \square}){\rm Im} (D) \oplus (1-{\rm P}_{\ker \square}){\rm Im}(D^\dagger)~.  
      \eea
        We can define the gauge fixing by restricting the fields to be in 
         ${\rm ker} (\square)  \oplus  (1-{\rm P}_{\ker \square}){\rm Im}(D^\dagger)$ where the fields in $\ker(\square)$ do not propagate; we actually   integrate over the fields in $(1-{\rm P}_{\ker \square}){\rm Im}(D^\dagger)$  which is Lagrangian in the space of BV fields. Indeed,
         modulo zero modes the properties 1 and 3  imply that the BV symplectic form  $\omega_{CS}$ restricted to ${\rm Im}(D^\dagger)$ is zero and
      it defines a good gauge fixing. The propagator will be 
      \bea
       K = \frac{D^\dagger}{\square} (1-P_{\ker\square})~,
      \eea
     which is the inverse of $D$ on the subspace $(1-P_{\ker \square}){\rm Im}(D^\dagger)$.  If additionally $D$ and $D^\dagger$ satisfy
 \begin{equation}\label{Hodges_like_relation}
 \ker D \cap {\rm Im} D^\dagger = \ker D^\dagger \cap {\rm Im} D =0~,
 \end{equation}
then it is easy to check that $\ker\square\cap{\rm Im}D=\ker\square\cap{\rm Im} D^\dagger=0$ so that we can remove the projection $P_{\ker\square}$ from (\ref{general_decomposition}) and $\ker\square=H(D)$, the cohomology of $D$. If ${\cal C}=\Omega^\bullet(\Sigma)$, with $\Sigma$ compact, and $D=d$, $D^\dagger=d^\dagger$, de Rham differential and codifferential, respectively, then the Hodge decomposition implies that  (\ref{Hodges_like_relation}) are satisfied. In general we will not assume (\ref{Hodges_like_relation}).

 Let us proceed to evaluate the action (\ref{BV-general-CS}) on $  {\rm Im} (D^\dagger)$.
  First of all we can define a pairing between fields
  \bea\label{pairing-formal}
   \langle {\cal A}^a, {\cal A}^b \rangle \equiv \int d\mu ~{\cal A}^a  \xi^b ~,
 \eea
  where $A^b = D^\dagger (\xi^b)$. This pairing is well-defined (so it does not depend on a choice of $\xi^a$), it is 
  non-degenerate on ${\rm Im} (D^\dagger)$ and it is symmetric. We use property 3 to establish these facts. 
  Let us introduce the anti-symmetric bracket
  \bea\label{def-general-bracket}
   \{ {\cal A}^a , {\cal A}^b \} \equiv D^\dagger ({\cal A}^a {\cal A}^b )~,
 \eea
  which is compatible with the paring $\langle~,~\rangle$
  \bea
    \langle {\cal A}^a, \{ {\cal A}^b , {\cal A}^c\} \rangle = - \langle \{ {\cal A}^b,  {\cal A}^a \} , {\cal A}^c  \rangle~.
  \eea
  Using these structures we can rewrite the action (\ref{BV-general-CS}) evaluated on ${\rm Im} (D^\dagger)$
    as follows
    \bea\label{GF-formal-action}
     S_{CS} = \frac{1}{2} \langle {\cal A}^a, \square {\cal A}^b \rangle \delta_{ab} +  \frac{1}{6}
     \langle {\cal A}^a, \{ {\cal A}^b , {\cal A}^c\} \rangle  f_{abc}~.
    \eea
     If the operator $D^\dagger$ is a second-order operator then the bracket (\ref{def-general-bracket}) is a Lie bracket and $\langle~,~\rangle$ is its invariant pairing. 
      Thus we observe that the gauge-fixed action of formal Chern-Simons theory has these additional algebraic structures which are not associated to
       the standard symmetries of the action.
A second order $D^\dagger$ defines a Gerstenhaber algebra structure on $\cal A$. The relation \eqref{basic-prop-2} implies that $({\cal A}, D, D^\dagger)$ forms a so-called BV${}^\Box$ algebra,  {\it i.e.} a BV algebra twisted by $\square$, according to \cite{Reiterer:2019dys}.
  
   To relate the above formal discussion to standard color-kinematics duality we work in $\mathbb{R}^d$, and further assume that the algebra $\mathcal{C}$ is a collection of differential forms possibly with some additional constraints.
   We require that $\square$ is the Laplace operator defined on this set of differential forms and it 
      never takes us outside of this set of forms. Once we assume this then the property (\ref{basic-prop-2}) becomes a very strong requirement on the operator $D^\dagger$. 
       In flat space $\mathbb{R}^d$ we can use the Fourier transform and the Laplace operator $\square$ corresponds to multiplication by $p^2$.
In addition, in flat space the zero modes do not cause any trouble since $\ker(\square)$ would correspond
    to poles in Feynman diagrams, and can be controlled with the Feynman $i\epsilon$ prescription.  To summarize:   if on $\mathbb{R}^d$ the formal Chern-Simons theory admits a second order 
    $D^\dagger$ operator  which satisfies properties (\ref{basic-prop-1}), (\ref{basic-prop-2}) with $\square$ being canonical Laplace operator, and  (\ref{basic-prop-3}) 
     then the gauge-fixed theory has off-shell color-kinematics duality.  In the rest of the paper we provide two examples which fit this formal construction and from now 
      on we assume that $\square$ is the canonical Laplace operator on the differential forms.

It is important to stress that in curved space it may not be possible to strip off the inverse Laplacian, and thus it is not straightforward to define kinematic numerators in general (although in specific spacetimes, for example symmetric spaces \cite{Diwakar:2021juk,Herderschee:2022ntr,Cheung:2022pdk}, realizations of color-kineamtics duality exist). In addition, on compact manifolds zero modes may appear in internal lines of the diagrams, requiring the use of $(1-P_{\textrm{ker}\square})$ and further complicating the standard interpretation of color-kinematics duality as arising from kinematic numerators that can be stripped off from diagrams. Despite these possible complications with defining numerators, the algebraic structures defined by $D^\dagger$ still persist. In both cases studied below we therefore present the kinematic algebras in curved space, as well as give examples in flat space.

 \section{4D BF-theory}\label{s:BF}
  
 As a first example of our formal construction we consider 4D BF-theory. Although it is topological we can still formally study off-shell amplitudes. The BF theory is defined by the following four-dimensional action
 \bea\label{BF-phys-1}
  S_{BF} = 2 \int {\rm Tr} \Big ( \phi d A + \phi A A \Big ) =
   \int  \Big (  \phi^a  d A^b \delta_{ab} + \frac{1}{2} \phi^a A^b A^c f_{abc} \Big )~,
 \eea
  where we assume that we have a compact Lie algebra $\mathfrak{g}$, $A$ is a connection one-form and $\phi$ is a two-form 
   which transforms in the adjoint with respect to the gauge group. The integral is over any smooth $4$-manifold $\Sigma_4$. All following discussions can be generalized to BF theory in arbitrary dimensions but for the sake of clarity we concentrate on the 4D example. 
 
 \subsection{BF-theory as a Chern-Simons theory}
 
 There exists a canonical construction of the BV master action for BF-theory which is the AKSZ construction \cite{Alexandrov:1995kv}.
  Consider the following supermaps
  \be
  T[1]\Sigma_4~\longrightarrow~\mathfrak{g}[1] \oplus \mathfrak{g}[2]~,
  \ee
   which correspond to two superfields
 \bea
 && \mathbf{A}^a = A^a_0 + A^a_1 + A^a_2 + A^a_3 + A^a_4~,\\
  && \mathbf{\Phi}^a = \phi^a_0 + \phi^a_1 + \phi^a_2 + \phi^a_3 + \phi^a_4~,
 \eea
  which are expanded in differential forms. 
  Here  $\mathbf{A}$ is of degree $1$ and $\mathbf{\Phi}$ is of degree $2$. 
   $A_1$ is a one-form of degree $0$ and $\phi_2$ is a two-form of degree $0$, so these are physical fields. When it is unambiguous we suppress the form degree on these two physical fields, $A^a=A^a_1$ for example. The BV master action is given by
    \be\label{BV-BF-2f}
     S_{BF} =  \int d^4 x d^4 \theta \Big ( \mathbf{\Phi}^a d \mathbf{A}^b \delta_{ab} + \frac{1}{2}
      \mathbf{\Phi}^a \mathbf{A}^b \mathbf{A}^c f_{abc} \Big )
    \ee
    with $d$ being de Rham operator on the differential forms. The corresponding BV symplectic structure is given by 
    \be\label{BV-symlp-2f}
     \omega_{BF} =  \int d^4 x d^4 \theta~\delta \mathbf{A}^a \wedge \delta \mathbf{\Phi}^b \delta_{ab}~.
    \ee
    If we expand the action (\ref{BV-BF-2f}) in components and set all fields to zero except physical fields
     then we arrive at the original action (\ref{BF-phys-1}).
    
   We will proceed to develop the underlying Frobenius algebra in detail. We can write our differential in the following complex
         \begin{equation}\label{BF_source}
\begin{tikzcd}
 & \Omega^0   \arrow[r,"d"] & \Omega^1  \arrow[r, "d"]  &  \Omega^{2}           \arrow[r, "d"]  &  \Omega^3   \arrow[r, "d"]   &\Omega^4 \\
  \Omega^0  \arrow[r, "d"]  & \Omega^1     \arrow[r, "d"]  & \Omega^{2}      \arrow[r, "d"]   &  \Omega^3   \arrow[r, "d"] &  \Omega^4     
     \end{tikzcd}
        \end{equation}
    where the first line corresponds to superfield $\mathbf{A}$ of degree $1$  and second line to superfield $\mathbf{\Phi}$ of degree $2$.
     If we introduce the additional odd coordinate $\zeta$ of degree $-1$ then one can combine these superfields in one single 
      superfield of degree $1$
      \be
        {\cal A} (x, \theta, \zeta) = \mathbf{A} (x, \theta) + \zeta \mathbf{\Phi} (x, \theta) ~,
      \ee
      which corresponds to maps
      \be
       T[1] \Sigma_4 \oplus \mathbb{R}[-1]~\longrightarrow~\mathfrak{g} [1]~.
      \ee
Thus the multiplication is dictated by the structure of the superfield ${\cal A}(x,\theta, \zeta)$, namely in (\ref{BF_source}) in the 
     first line the multiplication is given by the standard wedge product, in the second line the multiplication is trivial (since $\zeta^2=0$),
      and finally multiplication of an element in the first line by an element in the second line gives an element in the second line. Therefore the second line 
       is the module for the standard DGA of the first line.  The measure of degree $-3$ is given by the superintegration $d^4\theta~d\zeta$.
     Altogether we have the structure of  the differential graded Frobenius algebra $({\cal C}, D, \int d\mu)$ described in the previous section 
      with a pairing of the correct degree for formal Chern-Simons theory. 
        The diagram  (\ref{BF_source}) can be summarized as follows 
          \begin{equation}\label{BF_CFr}
\begin{tikzcd}
 & {\cal C}^{-1} \arrow[r,"d"] & {\cal C}^0  \arrow[r, "d"]  &  {\cal C}^1          \arrow[r, "d"]  &  {\cal C}^2   \arrow[r, "d"]   & {\cal C}^3     \arrow[r, "d"]   & {\cal C}^4
     \end{tikzcd}
        \end{equation}
     and we have the negative degrees subspace ${\cal C}^{-1}$ which corresponds to zero forms ($\phi_0$ component). 
        Using the formalism from the previous section we can rewrite the action (\ref{BV-BF-2f}) as follows
               \begin{equation}\label{BF_CS_form}
   S_{BF} = \int  d^4x d^4\theta d\zeta ~ \left (  \frac{1}{2} {\cal A}^a d   {\cal A}^b \delta_{ab}
    +  \frac{1}{6} {\cal A}^a {\cal A}^b {\cal A}^c f_{abc} \right ) \ ,
   \end{equation}
 and symplectic structure (\ref{BV-symlp-2f}) as  
   \bea\label{BF-fullBV-sympl}
   \omega_{BF} = \int  d^4x d^4\theta d\zeta ~ \delta {\cal A}^a \wedge \delta {\cal A}^b \delta_{ab} ~.
     \eea
   This is the Chern-Simons formulation of BF theory and in the next subsections we will discuss its gauge fixing and color-kinematics duality. 
     
  Let us mention that there exists another similar theory when obtained by deforming the differential $D= d + \partial_\zeta$ with the action
 \begin{equation}\label{DW_CS_form}
   S_{DW} = \int d^4x d^4\theta d\zeta ~ \left (  \frac{1}{2} {\cal A}^a (d + \frac{\partial}{\partial \zeta}) {\cal A}^b \delta_{ab}
    +  \frac{1}{6} {\cal A}^a {\cal A}^b {\cal A}^c f_{abc} \right )
   \end{equation}
   and the same BV symplectic form. If we restrict to the physical fields then the action would be
  \bea\label{DW-phys-1}
  S_{BF} = \int  \Big ( \phi^a  d A^b \delta_{ab} + \frac{1}{2} \phi^a \phi^b \delta_{ab} +  \frac{1}{2} \phi^a A^b A^c f_{abc} \Big )~.
 \eea
 This theory corresponds to the BV extension of Donaldson-Witten theory. However, as far as color-kinematics duality and diagramatics 
  are concerned, this theory is rather similar to BF theory (at least with the gauge fixing described below). 
   Thus we concentrate on BF theory only in what follows.

  \subsection{Gauge fixing and the kinematic Lie algebra}
  \label{ss:BF-gauge}
  
   Now let us proceed with the gauge fixing for this model by defining the operator $D^\dagger$ of degree $-1$ which satisfies 
    the properties (\ref{basic-prop-1})-(\ref{basic-prop-3}). 
Since our differential is $D=d$  we can use the most obvious choice for $D^\dagger=d^\dagger$, {\it i.e.} the codifferential defined by choosing a metric on $\Sigma_4$. It acts on the given dga as follows
        \begin{equation}\label{BF_gauge-fixing}
\begin{tikzcd}
 & \Omega^0  & \arrow[l,"d^\dagger"]  \Omega^1 & \arrow[l, "d^\dagger"]   \Omega^{2}      &     \arrow[l, "d^\dagger"]   \Omega^3 &  \arrow[l, "d^\dagger"]  \Omega^4 \\
  \Omega^0 & \arrow[l, "d^\dagger"]  \Omega^1  &   \arrow[l, "d^\dagger"] \Omega^{2}    &   \arrow[l, "d^\dagger"]   \Omega^3 &  \arrow[l, "d^\dagger"]  \Omega^4     
     \end{tikzcd}
        \end{equation}
Obviously we have that the properties (\ref{basic-prop-1}) and (\ref{basic-prop-2}) are satisfied, 
$\{D , D^\dagger \} = \{d , d^\dagger \}=\square$ and $(D^\dagger)^2 = (d^\dagger)^2=0$.  The property (\ref{basic-prop-3}) is also
 satisfied with the new measure 
 \be
   \int  d^4x d^4\theta d\zeta ~ d^\dagger (a) b = (-1)^{|a|}  \int  d^4x d^4\theta d\zeta ~ a d^\dagger b~. 
 \ee
  Thus if we evaluate the action (\ref{BF_CS_form}) on  ${\rm Im}(d^\dagger)$ we obtain the form (\ref{GF-formal-action}). 
   Let us rewrite the action  (\ref{GF-formal-action}) in terms of the $\mathbf{A}$ and $\mathbf{\Phi}$ superfields.
   The pairing is defined  in (\ref{pairing-formal}) and here it can be written as
    \be
  \langle {\cal A}^a , {\cal A}^b \rangle = \int  d^4x d^4 \theta d\zeta~ (\mathbf{A}^a +\zeta \mathbf{\Phi}^a ) (\xi_{A}^b + \zeta \xi_{\Phi}^b)
 \ee
   where ${\cal A}^b = d^\dagger (\xi_{A}^b + \zeta \xi_{\Phi}^b)$
 and  thus the only non-zero pairing is between $\mathbf{A}$ and $\mathbf{\Phi}$
\be\label{APhi-pairing-4D}
\langle \mathbf{\Phi}^a, \mathbf{A}^b \rangle = \int d^4x d^4 \theta~ \mathbf{\Phi}^a \xi_A^b = \int d^4x d^4 \theta~\mathbf{A}^b \xi_{\Phi}^a = \langle \mathbf{A}^b ,
 \mathbf{\Phi}^a \rangle~.
\ee 
 The bracket   defined in (\ref{def-general-bracket}) can be written for the superfields as follows,
 \bea 
 && \{ \mathbf{A}^a , \mathbf{A}^b \} = d^\dagger ( \mathbf{A}^a  \mathbf{A}^b )~,\nn\\
 &&  \{ \mathbf{A}^a , \mathbf{\Phi}^b \} = d^\dagger ( \mathbf{A}^a  \mathbf{\Phi}^b )~,\label{Lie-alg-BF}\\
 &&  \{ \mathbf{\Phi}^a , \mathbf{\Phi}^b \} = 0~,\nn
 \eea
  such that the pairing (\ref{APhi-pairing-4D}) is invariant with respects to these brackets. 
  Thus we can evaluate the action (\ref{BV-BF-2f}) on the gauge fixing
 \be\label{4D-BF-gf-super}
  S_{BF} = \langle \mathbf{\Phi}^a, \square \mathbf{A}^b \rangle \delta_{ab} + \frac{1}{2} \langle \mathbf{\Phi}^a, \{ \mathbf{A}^b , \mathbf{A}^c \} \rangle  f_{abc} ~,
 \ee
 which is exactly the same as the action (\ref{GF-formal-action}) after integrating $\zeta$ and using the above definitions of pairing and brackets.

Restricting the Lie algebra (\ref{Lie-alg-BF}) to the physical fields we can describe the kinematic Lie algebra on ${\rm Im}(d^\dagger)\subset\Omega^1 \oplus \Omega^2$ as follows 
   \bea
   && \{ A^a, A^b \}= d^\dagger (A^a A^b)~,\label{Lie-alg-1form-BF} \\
   && \{ A^a, \phi^b \} = d^\dagger ( A^a \phi^b )~, \label{Lie-alg-12-BF} \\
   && \{ \phi^a, \phi^b \}=0~. \label{Lie-alg-2-BF}
   \eea
    As we have explained in  Section \ref{s:proper}, using the metric we can map $A^a$ to divergenceless vector fields and $\phi^a$ to divergenceless bi-vector fields. The bracket (\ref{Lie-alg-1form-BF}) is then mapped in the bracket of the Lie algebra of divergenceless vector fields and  (\ref{Lie-alg-12-BF}) describes the action of a divergenceless vector field
       on a divergenceless bi-vector, given by the Lie derivative.   The kinematic Lie algebra is then the semi direct sum of the algebra of divergenceless vector fields with its representation on the space of divergenceless bi-vector fields. 

  As a consequence of the discussion in Section \ref{s:formal}, BF-theory on ${\mathbb R}^4$ exhibits the color-kinematics duality. In the next two subsections we expose this algebra in more detail in flat space.      
       
\subsection{The kinematic Lie algebra on ${\mathbb R}^4$}
 
We now consider $\Sigma_4 = {\mathbb R}^4$, equipped with the Euclidean metric.  We compute here the structure constants of the kinematic Lie algebra (\ref{Lie-alg-1form-BF},\ref{Lie-alg-12-BF},\ref{Lie-alg-2-BF}). 
    Introducing the Levi-Civita symbol $\epsilon_{\mu\nu\rho\sigma}$ for the Euclidean $\mathbb{R}^4$ with the following 
   normalization
 \bea
    \epsilon^{\mu\nu\rho\sigma} \epsilon_{\lambda\gamma \rho \sigma} = 2 (\delta^\mu_{\lambda} \delta^\nu_\gamma- \delta_\lambda^\nu \delta_\gamma^\mu )~,
   \eea 
 we have the following representation of the fields,
\bea
 &&  L^{\mu\nu} (p) = e^{ip \cdot x} \epsilon^{\mu\nu\rho\sigma} p_\rho \frac{\partial}{\partial x^\sigma} ~,\\
 && L^\mu (p) = e^{i p \cdot x} \epsilon^{\mu\nu\rho\sigma} p_{\nu} \frac{\partial}{\partial x^\rho}  \wedge \frac{\partial}{\partial x^\sigma} ~,
\eea
 where $L^{\mu\nu} (p)$ is the basis for divergenceless vector fields and $L^\mu (p)$ is the basis for divergenceless bi-vector fields. 
 Next we can calculate the structure constants explicitly in this basis
 \bea
 && [L^{\mu\nu} (p_1), L^{\rho\sigma}(p_2)] = F^{(\mu\nu)(\rho\sigma)}_{~~~~~~~~~(\delta\gamma)} (p_1, p_2; p_3) L^{\delta\gamma}(p_3)~,\\
 && [ L^{\mu\nu} (p_1), L^\rho (p_2) ] = F^{(\mu\nu)(\rho)}_{~~~~~~~(\sigma)} (p_1, p_2; p_3) L^\sigma(p_3)~,
 \eea
  where the first line is the standard Lie bracket and the second line is the Lie derivative for $L^{\mu\nu} (p)$ acting on $L^\mu (p)$.
  The structure constants are given by the following explicit expressions
  \bea
  &&F^{(\mu\nu)(\rho\sigma)}_{~~~~~~~~~(\delta\gamma)} (p_1, p_2; p_3) = \frac{i}{2} \epsilon^{\mu\nu\phi\zeta} p_{1\zeta } \epsilon^{\rho\sigma\xi \epsilon} p_{2 \epsilon} \epsilon_{ \phi \xi \delta \gamma} \delta (p_1 + p_2 - p_3)~,\\
 && F^{(\mu\nu)(\rho)}_{~~~~~~~(\sigma)} (p_1, p_2; p_3) = \frac{i}{2}\epsilon^{\mu\nu  \sigma_1 p_1}\epsilon^{\rho  \sigma_2\sigma_3 p_2}\epsilon_{\sigma_1\sigma_2\sigma_3 \sigma}\delta(p_1+p_2-p_3)~.
   \eea
     If we introduce the following notation 
     \be
     \Delta^{\mu\nu\rho} (p) =  \epsilon^{\mu\nu\rho\sigma} p_\sigma~,
    \ee
     then we have
    \begin{align*}
    &F^{(\mu\nu)(\rho\sigma)}_{~~~~~~~~~(\delta\gamma)} (p_1, p_2; p_3) = \frac{i}{2} \Delta^{\mu\nu\phi}(p_1)\Delta^{\rho\sigma\xi}(p_2) \epsilon_{ \phi \xi \delta \gamma} \delta (p_1 + p_2 - p_3)~,\\
 & F^{(\mu\nu)(\rho)}_{~~~~~~~(\sigma)} (p_1, p_2; p_3) = \frac{i}{2}\Delta^{\mu\nu\phi}(p_1) \Delta^{\rho \sigma_2\sigma_3}( p_2)\epsilon_{\sigma_1\sigma_2\sigma_3 \sigma}\delta(p_1+p_2-p_3)~.
	\end{align*}
  The Lie algebra satisfies by definition the Jacobi identities. In particular if we look at the brackets containing two vector fields and one-bivector 
   we get the following explicit form of the Jacobi identity	
 \bea\label{BF-Jacobi-structure}
  &&F^{(\mu\nu)(\phi)}_{~~~~~~~(\epsilon)} (p_1, p_4; p_5) F^{(\rho\sigma)(\gamma)}_{~~~~~~~(\phi)} (p_2, p_3; p_4)  
  -  F^{(\rho\sigma)(\phi)}_{~~~~~~~(\epsilon)} (p_2, p_4; p_5) F^{(\mu\nu)(\gamma)}_{~~~~~~~(\phi)} (p_1, p_3; p_4) \nn\\
  && =  F^{(\mu\nu)(\rho\sigma)}_{~~~~~~~~~(\delta\lambda)} (p_1, p_2; p_4) F^{(\delta\lambda)(\gamma)}_{~~~~~~~(\epsilon)} (p_4, p_3; p_5) ~.
 \eea    
   Moreover, the two sets of structure constants are related by 
    the following identity
\bea
F^{(\mu\nu)(\rho\sigma)}_{~~~~~~~~~(\delta\gamma)} (p_1, p_2; - p_3) \Delta^{\delta\gamma\epsilon} (-p_3) 
= F^{(\mu\nu)(\epsilon)}_{~~~~~~~(\gamma)} (p_1, p_3; -p_2) \Delta^{\gamma\rho\sigma} (-p_2)  \nn \\
= - i \Delta^{\mu\nu\alpha_1} (p_1) \Delta^{\rho\sigma\alpha_2}(p_2) \Delta^{\phi\alpha_3 \alpha_4} (p_3)  \epsilon_{\alpha_1 \alpha_2 \alpha_3 \alpha_4} \delta (p_1+p_2+ p_3)~.
\eea     
       
  \subsection{Tree diagramatics}   
  
  The structure constants introduced in the subsection above can be used to compute Feynman diagrams.  
  The tree diagrams in particular are controlled by the physical fields 
  \be\label{BF-phys-eval}
  S_{tree}=  \int  \Big (  \phi^a  d A^b \delta_{ab} + \frac{1}{2} \phi^a A^b A^c f_{abc} \Big )~, 
  \ee
   where we assume that $\phi$, $A$ are in ${\rm Im}(d^\dagger)$. Thus we can rewrite the gauge-fixed action as 
  \be
   S_{tree} = \langle \phi^a, \square A^b \rangle \delta_{ab} + \frac{1}{2} \langle \phi^a, \{ A^b , A^c \} \rangle f_{abc}~,
  \ee
 which is just restriction of  (\ref{4D-BF-gf-super}) to physical fields. Next we present the propagator and Feynman rules from this action, where the former is done using superfields. In momentum representation the superfield propagator
   is defined by the following conditions 
  \be
    d^\dagger  G(p, \theta, \zeta) =0~,~~~~d  G(p, \theta, \zeta)  = \theta^1 \theta^2 \theta^3 \theta^4 \zeta~,
  \ee
   where in the second relation on the right-hand side we have a delta function in the odd variables $\theta$ and $\zeta$. 
   These constraints can be solved with
\be
 G(p, \theta, \zeta) = \frac{1}{4!} \frac{d^\dagger (\epsilon_{\mu\nu\rho\sigma} \theta^\mu\theta^\nu \theta^\rho \theta^\sigma \zeta)}{p^2} ~ ,
\ee
 where  
 \be
  d^\dagger = - i p^\lambda \frac{\partial}{\partial \theta^\lambda} \ .
 \ee
 Using the notation 
 \be
  \Delta_{\mu\nu\rho} (p) =  \epsilon_{\mu\nu\rho\sigma} p^\sigma
 \ee
  the propagator can also be written as,
  \be
 G(p, \theta, \zeta) = \frac{i}{3!} \frac{\Delta_{\mu\nu\rho} (p) \theta^\mu \theta^\nu \theta^\rho \zeta}{p^2}~.
 \ee
  Finally, restricting to the physical fields we have the following propagator
  \be\label{BF-bos-prop}
   \langle \phi^a_{\mu\nu} (p) A^b_\rho (0) \rangle = i \frac{\Delta_{\mu\nu\rho} (p)}{p^2} \delta^{ab}~. 
  \ee
   Next we want to parametrize ${\rm Im} (d^\dagger)$ in the momentum representation as follows
 \bea
  A^a_\mu (x) = \frac{1}{(2\pi)^2} \int dp~ e^{ip\cdot x} \epsilon_{\mu}^{~\nu\rho\sigma} p_\nu \xi^a_{\rho\sigma} (p)
 \eea
  and 
    \bea
  \phi^a_{\mu\nu} (x) = \frac{1}{(2\pi)^2} \int dp~ e^{ip\cdot x} \epsilon_{\mu\nu}^{~~\rho\sigma} p_\rho \epsilon^a_{\sigma} (p)~,
 \eea
  where we assume that $p^\mu  \xi_{\mu\nu} (p) =0$ and $p^\mu \epsilon_\mu(p)=0$ to avoid extra ambiguities. 
   Up to irrelevant numerical factors, evaluating the kinetic term in (\ref{BF-phys-eval}) on these fields gives
   \bea
   \int \phi^a d A^b \delta_{ab}    \sim  \int d^4p ~ p^2 \Delta^{\mu\nu\rho}(p) \epsilon^a_{\mu}(-p) \xi^b_{\nu\rho}(p) \delta_{ab}~,
 \eea
   which produces the propagator (\ref{BF-bos-prop}).  Evaluating the interaction term  in (\ref{BF-phys-eval}) 
 gives us    
  \bea
  && \int \phi^a A^b  A^c f_{abc} \sim \\
  && \int d^4p_1 d^4p_2 d^4p_3~ 
    F^{(\mu\nu)(\rho\sigma)}_{~~~~~~~~~(\delta\gamma)} (p_1, p_2; -p_3)
   \Delta^{\delta\gamma\phi}(-p_3) \xi^b_{1\mu\nu} (p_1) \xi^c_{2\rho\sigma}(p_2) \epsilon^a_\phi (p_3) f_{abc}~,\nn
  \eea
   where we use the algebraic  notations from the previous subsection.
 If we look at the simplest diagram  (stripping off $1/p^2$ factors)
     \begin{align}
\begin{tikzpicture}[baseline=-0.2cm,scale=0.9]
\draw[thick] (-1,-1) -- (0,0) -- (-1,1);
\draw[thick] (0,0) -- (1.5,0)--(2.5,1);
\draw[thick] (1.5,0)--(2.5,-1);
\draw (-1.2,-1.3) node {$A^{a}(p_1)$};
\draw (-1.2,1.3) node {$A^{b}(p_2)$};
\draw (2.7,1.3) node {$A^{c}(p_3)$};
\draw (2.7,-1.3) node {$\phi^{d}(p_4)$};
\end{tikzpicture}
&= 
\langle \{ A^{a} (p_1), A^{b}(p_2) \}, \{ A^{c}(p_3), \phi^d(p_4) \} \rangle   \ ,
\end{align}
 and consider the cyclic sum in $(1,2,3)$ the result vanishes as it is identical to the Jacobi identity 
  (\ref{BF-Jacobi-structure}). This is the simplest explicit example of color-kinematics duality at tree level.
  
  More complicated tree diagrams can be constructed in a similar fashion. For example, for  the following diagram we can write kinematic numerator as
   \begin{eqnarray}
\begin{tikzpicture}[baseline={(0, 0cm)}]
\draw[thick] (-1,0) -- (1,0);
\draw[thick] (0,0) -- (0,0.5);
\draw[thick] (0.5,0) -- (0.5,0.5);
\draw[thick] (-0.5,0) -- (-0.5,0.5);
\node at (-1.25,0) {$A$};
\node at (-0.5,0.7) {$A$};
\node at (0,0.7) {$A$};
\node at (0.5,0.7) {$A$};
\node at (1.25,0) {$\phi$};
\end{tikzpicture} 
\, &\sim&
F^{(\mu_1\mu_2)(\mu_3\mu_4)}_{~~~~~~~~~~~~~(\nu_1\nu_2)}F^{(\nu_1\nu_2) (\mu_5\mu_6)}_{~~~~~~~~~~~~~(\rho_1\rho_2)}F^{(\rho_1\rho_2) (\mu_7 \mu_8)(\phi)}  \nonumber 
\end{eqnarray}
 where the last term is 
 \bea
 F^{(\rho_1\rho_2) (\mu_7 \mu_8)(\phi)} = F^{(\rho_1\rho_2) (\mu_7 \mu_8)}_{~~~~~~~~~~~~~(\sigma_1\sigma_2)} \Delta^{\sigma_1 \sigma_2 \phi}~.
 \eea
  Note that the Jacobi identities of this diagram will also require diagrams such as,
\bea
\begin{tikzpicture}[baseline={(0, 0cm)}]
\draw[thick] (-1,0) -- (1,0);
\draw[thick] (0,0) -- (0,0.5);
\draw[thick] (0.5,0) -- (0.5,0.5);
\draw[thick] (-0.5,0) -- (-0.5,0.5);
\node at (-1.25,0) {$A$};
\node at (-0.5,0.7) {$A$};
\node at (0,0.7) {$\phi$};
\node at (0.5,0.7) {$A$};
\node at (1.25,0) {$A$};
\end{tikzpicture} 
\, &\sim&
F^{(\mu_1\mu_2)(\mu_2\mu_3)}{}_{(\nu_1\nu_2)}F^{(\nu_1\nu_2)(\phi)}{}_{(\beta)}F^{(\beta)(\mu_5\mu_6)(\mu_7\mu_8)} .
\eea

 \section{Yang-Mills theory in 2D}\label{s:YM}
 
 In this section we present an action realizing off-shell color-kinematics duality for 2D Yang-Mills theory. We start with the following Euclidean action defined on a smooth manifold $\Sigma_2$
 \be\label{simple-2DYM-action}
  S^{(0)}_{YM} = - \frac{1}{2} \int F^a \wedge \star F^b \delta_{ab}~,
 \ee
  where the field strength $F$ is defined by,
  \be
    F^a = dA^a + \frac{1}{2} f^a_{~bc} A^b A^c~,
  \ee
   and we use the same notation for Lie algebra data as before. We can introduce the auxiliary scalar field $\phi^a$ in the adjoint representation 
    of Lie algebra $\mathfrak{g}$ and define the following action 
\be\label{YM-phys}
  S^{(0)}_{YM} = \int \Big ( \phi^a dA^b \delta_{ab} + \frac{1}{2} \phi^a \star \phi^b \delta_{ab} + \frac{1}{2} \phi^a A^b A^c f_{abc} \Big )~.
 \ee
  Integrating out $\phi^a$ we get the Yang-Mills action
(\ref{simple-2DYM-action}).  Here we assume that $\star 1 = {\rm vol}$ is the volume form.

 \subsection{2D YM as CS theory}
 
 We start by presenting the BV extension of the theory defined by the action (\ref{YM-phys}).  2D Yang-Mills theory can be realized as a deformation of 
  2D BF-theory, thus we start from the AKSZ construction of the latter.  The space of supermaps is defined as
  \be\label{2D-space-fields-1}
   T[1]\Sigma_2~\longrightarrow~\mathfrak{g}[1] \oplus \mathfrak{g}[0]~,
  \ee
 which corresponds to   two superfields  
  \bea
 && \mathbf{A}^a = A^a_0 + A^a_1 + A^a_2  ~,\\
  && \mathbf{\Phi}^a = \phi^a_0 + \phi^a_1 + \phi^a_2  ~,
 \eea
  which are expanded in differential forms as before. 
  Here  $\mathbf{A}$ is of degree $1$ and $\mathbf{\Phi}$ is of degree $0$. 
 When it is unambiguous we suppress the form degree on $\phi^a\equiv \phi^a_0$ and $A^a\equiv A^a_1$, which are the physical fields. 
  The odd symplectic form is defined as follows
  \be\label{2D-symplec-1}
 \omega_{2D} =  \int  d^2x d^2\theta~  
    \delta {\bf \Phi}^a \wedge \delta {\bf A}^b \delta_{ab} ~,
 \ee
  and the standard  BV action for 2D BF-theory is
  \bea
   S_{BF} =  \int  d^2x d^2\theta~ \Big ( \mathbf{\Phi}^a d \mathbf{A}^b \delta_{ab} + \frac{1}{2} \mathbf{\Phi}^a \mathbf{A}^b \mathbf{A}^c f_{abc} \Big )~.
  \eea
   The BV formulation of 2D Yang-Mills theory corresponds to the following deformation of the above action 
    \bea\label{2D-BV-action}
   S_{2DYM} =  \int  d^2x d^2\theta~ \Big ( \mathbf{\Phi}^a d \mathbf{A}^b \delta_{ab} + \frac{1}{2} {\rm vol}~ \mathbf{\Phi}^a \mathbf{\Phi}^b \delta_{ab}   + \frac{1}{2} \mathbf{\Phi}^a \mathbf{A}^b \mathbf{A}^c f_{abc} \Big )~,
  \eea
  which satisfies the master equation with respect to the symplectic form (\ref{2D-symplec-1}).
 This deformed action can be recast 
   in the form of a generalized Chern-Simons through the deformation of the de Rham differential on $T[1] \Sigma \times \mathbb{R}[1]$, where we introduce an auxiliary odd variable $\zeta$ of degree $1$.   
    The space of fields (\ref{2D-space-fields-1}) can be alternatively rewritten as
    \be
     T[1] \Sigma \times \mathbb{R}[1]~\longrightarrow~\mathfrak{g}[1]~, 
    \ee
    where now we define a new superfield of degree $1$
  \be
  {\cal A} (x, \theta, \zeta) = {\bf A}(x, \theta) + \zeta {\bf \Phi}(x, \theta)~.
  \ee
   The space $T[1] \Sigma \times \mathbb{R}[1]$ is equipped with the following differential 
\be
 D = d + {\rm vol} \frac{\partial}{\partial \zeta}  = \theta^\mu \frac{\partial}{\partial x^\mu} + {\rm vol} \frac{\partial}{\partial \zeta}~,
\ee
 which is clearly a deformation of the de Rham differential. Finally, this DGA can be encoded in the following diagram 
   \begin{equation}\label{2D-YM-complex}
\begin{tikzcd}
\Omega^0 (\Sigma_2) \arrow[r,"d"] & \Omega^1 (\Sigma_2) \arrow[r,"d"] & \Omega^{2}(\Sigma_2)      &     \\
& \zeta \Omega^{0} (\Sigma_2)  \arrow[r,"d"] \arrow[ur,"{\rm vol}\partial_\zeta"]   & \zeta \Omega^1 (\Sigma_2)  \arrow[r,"d"] & \zeta \Omega^2(\Sigma_2)    
\end{tikzcd} \ .
\end{equation}
The field space is equipped with the pairing of degree $-3$
  given by the integral $\int  d^2x d^2\theta d\zeta$ which is compatible with $D$  as described in Section \ref{s:formal}.
      Thus we can recast the BV formulation of 2D Yang-Mills theory as a formal Chern-Simons theory with the following superfield action 
 \be\label{super-2D-CS}
 S_{YM} = \int d^2 x d^2\theta  d\zeta ~ \Big ( \frac{1}{2}{\cal A}^a D {\cal A}^b \delta_{ab} + \frac{1}{6}{\cal A}^a{\cal A}^b{\cal A}^c f_{abc} \Big )~,
 \ee
 with the corresponding odd  symplectic structure,
\bea\label{YM2D-fullBV-sympl}
   \omega_{YM} = \frac{1}{2} \int  d^2x d^2\theta d\zeta ~ \delta {\cal A}^a \wedge \delta {\cal A}^b \delta_{ab} ~.
   \eea
  If in the Chern-Simons action (\ref{super-2D-CS}) we integrate over the odd $\zeta$ variable then we end up with the action (\ref{2D-BV-action}), and similarly for the symplectic form (\ref{YM2D-fullBV-sympl}) and (\ref{2D-symplec-1}). Next we have to look for a suitable gauge fixing.

 \subsection{In search of gauge fixing} 

Within our framework the gauge fixing is defined by suitable operator $D^\dagger$ of degree $-1$.
Thus  in this subsection we present the derivation of $D^\dagger$ which satisfies the properties (\ref{basic-prop-1})-(\ref{basic-prop-3}), we provide a detailed analysis to illustrate that the gauge-fixing operator we find is rather unique given the restrictions we make on the total number of derivatives it carries.

Let us start by  repacking  the complex (\ref{2D-YM-complex}) as follows
  \bea
 \left ( \begin{array}{c}
  \Omega^0 \\
  0
  \end{array} \right ) \underrightarrow{D_1}
   \left ( \begin{array}{c}
  \Omega^1 \\
 \zeta \Omega^0
  \end{array} \right )  \underrightarrow{D_2}
     \left ( \begin{array}{c}
  \Omega^2 \\
  \zeta \Omega^1
  \end{array} \right )   \underrightarrow{D_3}
     \left ( \begin{array}{c}
  0 \\
  \zeta \Omega^2
  \end{array} \right )  
  \eea
  with the differential written in the following matrix form
  \bea
   D_1 =  \left ( \begin{array}{cc}
   d& 0 \\
  0 & 0
  \end{array} \right ) ~,~~~~~
  D_2 =  \left ( \begin{array}{cc}
   d& \star \partial_\zeta \\
  0 & d
  \end{array} \right ) ~,~~~~~D_3 =  \left ( \begin{array}{cc}
   0& 0 \\
  0 & d
  \end{array} \right ) ~,
  \eea
   where $\partial_\zeta = \frac{\partial}{\partial \zeta}$ and here the action of $\star$ on zero forms is the same as multiplication by a volume form. 
    These operators satisfy the property
   \bea
     D_2 D_1 =0~,~~~~~D_3 D_2=0~,
   \eea
    as required for $D$ to be nilpotent.

Next we construct an ansatz for the operator $D^\dagger$ of degree $-1$ which acts as follows
    \bea
 \left ( \begin{array}{c}
  \Omega^0 \\
  0
  \end{array} \right ) \underleftarrow{D^\dagger_1}
   \left ( \begin{array}{c}
  \Omega^1 \\
  \zeta \Omega^0
  \end{array} \right )  \underleftarrow{D^\dagger_2}
     \left ( \begin{array}{c}
  \Omega^2 \\
 \zeta \Omega^1
  \end{array} \right )   \underleftarrow{D^\dagger_3}
     \left ( \begin{array}{c}
  0 \\
  \zeta \Omega^2
  \end{array} \right )  
  \ .
  \eea
We assume that the ansatz for $D^\dagger$ is built from $d$, $d^\dagger=-\star d\star$, and $\star$. In order to avoid introducing spurious poles we also restrict the operator to be at most first order in spatial derivatives, meaning it is zeroth order or linear in $d$ and $d^\dagger$. Our ansatz takes the form,
   \bea
   && D^\dagger_1 =  \left ( \begin{array}{cc}
   d^\dagger + \alpha_1 \star d & \gamma_1  \partial_\zeta\\
  0 & 0
  \end{array} \right ) ~,~~
  D^\dagger _2 =  \left ( \begin{array}{cc}
   d^\dagger + \alpha_2 d \star & ( \star \gamma_2 + \gamma_3)\partial_\zeta \\
  0 & d^\dagger + \beta_1 \star d
  \end{array} \right ) ~,\nn \\
&&  D^\dagger_3 =  \left ( \begin{array}{cc}
   0& \gamma_4 \partial_\zeta \\
  0 & d^\dagger + \beta_2 d \star
  \end{array} \right ) ~,
  \eea
   with $\alpha_i$, $\beta_i$, $\gamma_i$ some constants to be fixed. 
 Note that each entry in the matrix representation of $D^\dagger$ is at most second order, which is required (but not neccesarily sufficient) for the operator to be second order overall.

  We start from property (\ref{basic-prop-1}) which is written as
 \be
  D^\dagger_2 D^\dagger_3=0~,~~~~D^\dagger_1 D^\dagger_2 =0~,
 \ee
  and  we get the following conditions
  \bea
 && \beta_1+ \beta_2 =0~,\nn \\
  && \gamma_4 + \beta_2 \gamma_2 - \gamma_3 =0~,  \label{result-1A}\\
  && \alpha_2 \gamma_4 - \gamma_2 - \gamma_3 \beta_2 =0~,\nn
 \eea
  and 
   \bea
  && \alpha_1 + \alpha_2 =0~,\nn \\
 &&  \gamma_3 - \alpha_1 \gamma_2 - \gamma_1 =0~, \label{result-1B}\\
  && \gamma_2 + \alpha_1 \gamma_3 - \gamma_1 \beta_1 =0~, \nn 
 \eea
  correspondingly.  Note that $ \star^2 = (-1)^k$ when acting on $k$-forms, see Appendix \ref{s:2dconventions} for further details on conventions.
  
  Next we require property  (\ref{basic-prop-2}) which is written as follows in our notation
 \bea 
 && D^\dagger_1 D_1 = \left ( \begin{array}{cc}
 \square & 0\\
 0 & 0
 \end{array} \right )~, \label{cond-2YM-A} \\
&& D_1 D^\dagger_1 + D^\dagger_2 D_2 = \left ( \begin{array}{cc}
 \square & 0\\
 0 & \square
 \end{array} \right )~,\label{cond-2YM-B}\\
&& D_2 D^\dagger_2 + D^\dagger_3 D_3 = \left ( \begin{array}{cc}
 \square & 0\\
 0 & \square
 \end{array} \right )~, \label{cond-2YM-C}\\
 &&  D_3 D^\dagger_3 = \left ( \begin{array}{cc}
 0 & 0\\
 0 & \square
 \end{array} \right )~, \label{cond-2YM-D}
\eea
 where on the right-hand side we have the Laplace operators acting diagonally on our complex. The conditions
  (\ref{cond-2YM-A}) and (\ref{cond-2YM-D}) do not imply any restrictions on the constants. 
   The condition (\ref{cond-2YM-B}) implies the following constraints
\bea
&& \alpha_1 + \alpha_2 =0~,\nn\\
 && \gamma_2= -1~, \label{result-2A}\\
 && \gamma_1 +  \alpha_2 - \gamma_3=0~,\nn
\eea
 and the condition (\ref{cond-2YM-C}) implies
\bea
&& \beta_1 + \beta_2 =0 ~,\nn \\
&&\gamma_2= - 1~, \label{result-2B} \\
&& \gamma_3 - \beta_1 - \gamma_4=0~.\nn
\eea
  Finally, we require  property  (\ref{basic-prop-3}) which in our conventions corresponds to two relations. 
   First is the relation
  \be
    \int d^2 \theta d \zeta~ D^\dagger_1 (\omega_1 + \zeta \omega_0) \zeta \omega_2 = 
    - \int d^2 \theta d \zeta~ (\omega_1 + \zeta \omega_0) D_3^\dagger (\zeta  \omega_2)~,
   \ee
 which implies 
 \be
  \gamma_1 = - \gamma_4~,~~~~~\alpha_1 = - \beta_2~.\label{result-3A}
 \ee
  The second relation is the following 
  \be
  \int d^2 \theta d \zeta~  D_2^\dagger (\omega_2 + \zeta \omega_1) (\tilde{\omega}_2 +  \zeta \tilde{\omega}_0) =   \int d^2 \theta d \zeta~ 
    (\omega_2 + \zeta  \omega_1) D_2^\dagger (\tilde{\omega}_2 +  \zeta \tilde{\omega}_1)~,
 \ee
  which implies 
  \be
   \gamma_3=0~,~~~~~\alpha_2 = - \beta_1~. \label{result-3B}
  \ee
  Here $\omega_i$ and $\tilde{\omega}_i$ are differential forms of degree $i$ and we have used the following identities for the differential forms
  \be
   \int (d^\dagger \omega_1) \omega_2 = - \int \omega_1 d^\dagger \omega_2~,~~~~~\int (\star d \omega_1) \omega_2 = \int \omega_1 (d\star \omega_2)~.
  \ee
    Combining all requirements 
  (\ref{result-1A}),  (\ref{result-1B}),  (\ref{result-2A}),  (\ref{result-2B}),  (\ref{result-3A}) and  (\ref{result-3B}) for the constants we get 
   only two possible solutions
  \be
   \gamma_2=-1~,~~~\gamma_1 = - \gamma_4 = \pm i ~,~~~\alpha_1=\beta_1 = - \alpha_2 = - \beta_2= \pm i~,~~~~\gamma_3=0~. 
  \ee
   If we choose the first solution and choose a complex structure compatible with the metric ({\it i.e.} a K\"ahler structure), then the operator $D^\dagger$ can be encoded in the following diagram (see Appendix \ref{s:2dconventions} for conventions)
    \begin{equation}
\begin{tikzcd}[column sep=huge]
\Omega^0 (\Sigma_2) & \Omega^1 (\Sigma_2) \arrow[l,"2\partial^\dagger"] & \Omega^{2} (\Sigma_2) \arrow[l,"2\partial^\dagger"]      &     \\
& \arrow[ul,"i\partial_\zeta"]\zeta\Omega^{0}  (\Sigma_2) &  \arrow[ul,"-\star\partial_\zeta"] \arrow[l,"2\partial^\dagger"] \zeta \Omega^1 (\Sigma_2)
  & \arrow[l," 2\partial^\dagger"] \arrow[ul,"-i \partial_\zeta"] \zeta \Omega^2 (\Sigma_2)
\end{tikzcd}
\end{equation}
 with the other solution given by complex conjugation of this solution. We see that $D^\dagger$ is a complex operator in Euclidean space
  and thus it would formally require the complexification of the fields. In Minkowski space the analogous operator will be real, to obtain it we need to perform a Wick rotation for the even coordinate $y \rightarrow i y$, as well as for the odd coordinates
$\theta_y\rightarrow i \theta_y$ and    $\zeta\rightarrow i \zeta$,  and all algebra goes through with real fields. 
   
Although the diagram does not make this obvious, $D^\dagger$ is  a second-order operator. To make this manifest one can write it as,
\begin{equation}\label{codiff_YM2}
D^\dagger = 2\partial^\dagger + \mathcal{I} \frac{\partial}{\partial\zeta} \ ,
\end{equation}
  where in a local basis $\mathcal{I}=i(\iota_{d\bar{z} }d\bar{z} - d\bar{z} \iota_{d\bar{z} })$ satisfies $\mathcal{I}^2=-1$ and $[\partial^\dagger,\mathcal{I}]=0$, and is a first-order operator. Thus, in analogy with equation \eqref{eqn:oneformsbracket} for $d^\dagger$, our $D^\dagger$  here defines a Lie bracket on the space of fields,
  \begin{equation}
  \{\mathcal{A}^a, \mathcal{A}^b\} = D^\dagger (\mathcal{A}^a \mathcal{A}^b) - D^\dagger \mathcal{A}^a \mathcal{A}^b + \mathcal{A}^a D^\dagger \mathcal{A}^b \ ,
  \end{equation} 
 containing a Lie subalgebra when restricting to fields in ${\rm Im}(D^\dagger)$.

 \subsection{The kinematic Lie algebra}

  Due to the structure of the operator $D^\dagger$ one can show that $\ker (D^\dagger)$ coincides with ${\rm Im} (D^\dagger)$. The kernel of 
   this operator is defined as follow
   \be
    D^\dagger {\cal A} =0~\longleftrightarrow~ 2\partial^\dagger \mathbf{A} + {\cal I} \mathbf{\Phi} =0~,
   \ee
 where the condition $\partial^\dagger \mathbf{\Phi}=0$ follows from above since $[\partial^\dagger, {\cal I}]=0$ and ${\cal I}$ is an invertible operator with ${\cal I}^{-1}=-{\cal I}$.
   The image of $D^\dagger$ is defined as 
   \be
    {\cal A} = D^\dagger \xi = D^\dagger (\xi_{A} + \zeta \xi_{\Phi})~ \longleftrightarrow~ \mathbf{A} = 2\partial^\dagger \xi_A + {\cal I} \xi_\Phi~,~~~~\mathbf{\Phi} = - 2\partial^\dagger \xi_\Phi \ .
   \ee
   Since $(D^\dagger)^2=0$ we have ${\rm Im} (D^\dagger) \subset \ker (D^\dagger)$. But here since ${\cal I}$ is an invertible operator it follows that 
    $\ker (D^\dagger) \subset    {\rm Im} (D^\dagger)$. 
     Modulo  zero modes,    the gauge fixing is defined by the conditions
     \be\label{2D-gaugef-cond}
      \mathbf{\Phi} = 2 {\cal I} \partial^\dagger \mathbf{A}~,
     \ee
     for superfields, or in components
     \be
      \phi_0 = 2i \partial^\dagger A_1~,~~~~\phi_1 = -2 \star \partial^\dagger A_2~,~~~~\phi_2=0~,
     \ee
  and the symplectic form \eqref{YM2D-fullBV-sympl} vanishes on these conditions.
 
  As before, we can define the pairing on ${\rm Im} (D^\dagger)$
  \bea\label{2D-pairing-general}
  \langle {\cal A}^a, {\cal A}^b \rangle &&  = \int d^2x d^2 \theta d \zeta~ {\cal A}^a {\xi}^b =  \int d^2x d^2 \theta  ~\mathbf{A}^a {\cal I}\mathbf{A}^b \nn \\
&&   = \int \Big (i A^a_0 A^b_2 -  A^a_1 \star A_1^b  - i A_2^a A_0^b \Big) \equiv  \langle \mathbf{A}^a, \mathbf{A}^b \rangle ~,\label{2D-def-pairing}
  \eea
  where we have used that ${\cal A}^b$ is in ${\rm Im} (D^\dagger)$, namely ${\cal A}^b = \mathbf{A}^a + 2 \zeta {\cal I} \partial^\dagger \mathbf{A}^a$.
     Note that this pairing is symmetric due to the parity of the different fields.
Proceeding to work on  ${\rm Im} (D^\dagger)$ we define the following Lie bracket
   \begin{align}
   \label{2dactiongaugefixed}
     \{ {\cal A}^a, {\cal A}^b \} = 
    D^\dagger ({\cal A}^a {\cal A}^b)  &= 2 \partial^\dagger (\mathbf{A}^a \mathbf{A}^b) + 2 {\cal I}\Big ( ({\cal I} \partial^\dagger \mathbf{A}^a) \mathbf{A}^b -
    \mathbf{A}^a ({\cal I}\partial^\dagger \mathbf{A}^b ) \Big ) \nn \\ 
    &- 2\zeta \partial^\dagger \Big ( ({\cal I} \partial^\dagger \mathbf{A}^a) \mathbf{A}^b - \mathbf{A}^a ({\cal I} \partial^\dagger \mathbf{A}^b ) \Big )  \ ,
   \end{align}
and evaluating the action (\ref{2D-BV-action}) on the gauge-fixing condition (\ref{2D-gaugef-cond}) we get the following gauge-fixed action
 \be\label{2D-YM-gauged}
  S = \frac{1}{2} \langle \mathbf{A}^a , \square \mathbf{A}^b \rangle \delta_{ab} + \frac{1}{6} \langle \mathbf{A}^a , \{ \mathbf{A}^b , \mathbf{A}^c \} \rangle f_{abc}\ ,
 \ee
  where the pairing is defined in equation \eqref{2D-pairing-general} and the bracket is,
     \begin{align}
     \{ {\mathbf A}^a, {\mathbf A}^b \} = 2 \partial^\dagger (\mathbf{A}^a \mathbf{A}^b) + 2 {\cal I}\Big ( ({\cal I} \partial^\dagger \mathbf{A}^a) \mathbf{A}^b -
    \mathbf{A}^a ({\cal I}\partial^\dagger \mathbf{A}^b ) \Big )  \ .
   \end{align}
   In components this bracket on the physical sector gives,
   \bea
&&    \{ A^{(1,0)a},  A^{(1,0)b} \} = -2 \Big (  (\partial^\dagger A^{(1,0)a})  A^{(1,0)b}   - A^{(1,0)a} (\partial^\dagger  A^{(1,0)b}) \Big ) ~,\label{2D-bracket1}\\
 &&   \{ A^{(1,0)a},  A^{(0,1)b} \} = 2 \partial^\dagger ( A^{(1,0)a}  A^{(0,1)b} )  + 2 (\partial^\dagger A^{(1,0)a})  A^{(0,1)b} ~,\label{2D-bracket2}\\
&&    \{ A^{(0,1)a},  A^{(0,1)b} \} =0~.\label{2D-bracket3}
  \eea
 Note that the pairing and the brackets are compatible
     \be\label{pairin-inv-phys}
   \langle  \{ A^{(1,0)b},  A^{(1,0)a} \}, A^{(0,1)c} \rangle = - \langle  A^{(1,0)a},  \{ A^{(1,0)b} , A^{(0,1)c} \} \rangle~. 
   \ee 

Several alternative reinterpretations of this algebra are possible. For example, by using the metric to raise the index on ($1,0$)-forms to ($0,1$) vector fields, the kinematic Lie algebra can then be described as the semidirect sum of the Lie algebra of ($0,1$) vector fields together with its representation on ($0,1$)-forms given by 
\begin{equation}
[v^{(0,1)}, \nu^{(0,1)}] = 2 {\bar L}_{v^{(0,1)}}(\nu^{(0,1)}) + 2 {\rm div}(v^{(0,1)}) \nu^{(0,1)}
\end{equation}
where $v^{(0,1)}$ is an element of the section of $T^{(0,1)}\Sigma_2$ and $\nu^{(0,1)}$ is an element of the secontion of $T^*{}^{(0,1)}\Sigma_2$ and ${\bar L}_v=\iota_v\bar{\partial}+{\bar\partial}\iota_v$. 
   
  So far we have ignored the issue of zero modes, that does not create  problems in flat $\mathbb{R}^2$, see next subsection. However, on 
   compact spaces we need to analyse the issue of zero modes in the action (\ref{2D-YM-gauged}) since there will be a harmonic 
    component of $\mathbf{A}^a$ which is absent from the kinetic term but appears in the interactions. In addition, an analogue of the Hodge decomposition does not necessarily work for $D$ and $D^\dagger$, therefore zero modes could appear in internal lines of diagrams. Since the structure and appearance of zero modes is quite specific to the manifold in question we leave further investigations of this topic to future work.  
 
  \subsection{Tree and loop diagrams}
  
To gain intuition for the kinematic algebra and make contact with previous sections we will compute explicit numerators in $\mathbb{R}^2$. 
Using the conventions from the appendix we have the following explicit superfield form 
 \begin{align}
D &=
\theta^z\frac{\partial}{\partial z} 
+ \theta^{\bar{z}}\frac{\partial}{\partial\bar{z}} 
+ \frac{i}{2}\frac{\partial}{\partial\zeta} \theta^z \theta^{\bar{z}} ~,\\
D^\dagger &= -4 \frac{\partial}{\partial \theta^{z}} \frac{\partial}{\partial \bar{z}} + \Big(i\frac{\partial}{\partial\theta^{\bar{z}}} \theta^{\bar{z}} - i\theta^{\bar{z}} \frac{\partial}{\partial\theta^{\bar{z}}}  \Big)\frac{\partial}{\partial\zeta}~.
\end{align}   

We start with example diagrams at tree level, and show also that in this theory the only non-trivial diagrams are at tree- and one-loop level, all higher loops vanish. Finally we also present a one-loop example of a Jacobi identity.
To begin, the gauge-fixed action for the physical fields,
  which is the same as evaluating the physical action (\ref{YM-phys}) on   $2 \partial^\dagger A^a + i \phi^a=0$, is
 \be\label{2D-tree-action}
  S_{tree} = \int \Big ( -  A^{(1,0)a} \star \square A^{(0,1)b} \delta_{ab}  + 2 i  (\partial^\dagger A^{(1,0)a})
   A^{(1,0)b} A^{(0,1)c} f_{abc} \Big )~.
 \ee
   Using   the definition of  the pairing (\ref{2D-pairing-general}) restricted to the physical fields 
   \be
  \langle A^{(1,0)a},  A^{(0,1)b} \rangle = - \int   A^{(1,0)a}   \star A^{(0,1)b} 
 \ee
 the action can be rewritten as 
  \be
  S_{tree} =  \langle A^{(1,0)a},  \square A^{(0,1)b} \rangle  \delta_{ab} + \frac{2}{3}  \langle  A^{(1,0)a},\{  A^{(1,0)b} , A^{(0,1)c} \}  \rangle f_{abc}~,
  \ee
 where the brackets  are defined by \eqref{2D-bracket1}-\eqref{2D-bracket3}.
We introduce the following expression for the vector fields in momentum representation 
 \be
  A^a_i (z, \bar{z}) = \int d^2p~ e^{\frac{i}{2} (p\bar{z} + \bar{p}z)}  \xi^a (p) \ ,
 \ee
such that the interaction terms in (\ref{2D-tree-action})  give rise to the vertex,
\begin{eqnarray}
\begin{tikzpicture}[baseline={(0, -0.1cm)}]
\draw[thick] (0,0.75) -- (0,0);
\draw[thick,stealth reversed-] (0,0.75/2) -- (0,0);
\draw[thick] (0,0) -- (0.87*0.75,-0.5*0.75);
\draw[thick,-stealth] (0,0) -- (0.87*0.75/2,-0.5*0.75/2);
\draw[thick] (0,0) -- (-0.87*0.75,-0.5*0.75);
\draw[thick,-stealth reversed] (0,0) -- (-0.87*0.75/2,-0.5*0.75/2);
\node at (0,0.9) {$p_1$};
\node at (-0.87*1.1,-0.5*1.1) {$p_3$};
\node at (0.87*1.1,-0.5*1.1) {$p_2$};
\end{tikzpicture}
=
2i (p_1 - p_2) \delta^2 (p_1 + p_2 + p_3) 
\end{eqnarray}
for the kinematic part of the interaction only. 
 Similarly, for the ghosts we find the vertex,
\begin{eqnarray}
\begin{tikzpicture}[baseline={(0, -0.1cm)}]
\draw[thick] (0,0.75) -- (0,0);
\draw[thick,stealth reversed-] (0,0.75/2) -- (0,0);
\draw[thick,dotted] (0,0) -- (0.87*0.75,-0.5*0.75);
\draw[thick,dotted] (0,0) -- (-0.87*0.75,-0.5*0.75);
\draw[thick,-stealth,dotted] (0,0) -- (0.87*0.75/2,-0.5*0.75/2);
\draw[thick,-stealth reversed,dotted] (0,0) -- (-0.87*0.75/2,-0.5*0.75/2);
\node at (0,0.9) {$p_1$};
\node at (-0.87*1.1,-0.5*1.1) {$p_3$};
\node at (0.87*1.1,-0.5*1.1) {$p_2$};
\end{tikzpicture} 
=
2i p_2 \delta^2(p_1+p_2+p_3) \ .
\end{eqnarray}
In these Feynman rules the incoming arrows indicate an incoming $(1,0)$ form, while outgoing arrows indicate $(0,1)$ forms, and similarly the ghost vertex also has an incoming arrow for the $(0,0)$ form and outgoing for $(1,1)$. Since the propagators preserve the direction of the arrows, we learn from this that, just like for the self-dual sector of Yang-Mills theory, the only non-trivial diagrams one can construct are at tree level or one loop \cite{Monteiro:2011pc}. Moreover, at one loop the diagrams consist of only incoming $(1,0)$ forms.
To compute diagrams we of course also require the propagator for the physical fields,
 \be\label{2D-YM-bos-prop}
  \langle A^{(1,0)a}  (p, \bar{p})  A^{(0,1)b} (0) \rangle = \frac{1}{2p\bar{p}} \delta^{ab}~.
 \ee
Using these Feynman rules, the off-shell four-point diagram can be written as,
 \begin{align}
\begin{tikzpicture}[baseline=-0.2cm,scale=0.9]
\draw[thick] (-1,-1) -- (0,0) -- (-1,1);
\draw[thick] (0,0) -- (1.5,0)--(2.5,1);
\draw[thick] (1.5,0)--(2.5,-1);
\draw[thick,-stealth] (-1,-1) -- (-0.5,-0.5);
\draw[thick,-stealth] (-1,1) -- (-0.5,0.5);
\draw[thick,-stealth] (2.5,1) -- (2,0.5);
\draw[thick,-stealth] (1.5,0)--(2.15,-0.65);
\draw[thick,-stealth] (0,0)--(0.85,0);
\draw[thick] (0.75,0)--(1.5,0);
\draw (-1.2,-1.3) node {$A^{z,a}(p_1)$};
\draw (-1.2,1.3) node {$A^{z,b}(p_2)$};
\draw (2.7,1.3) node {$A^{z,c}(p_3)$};
\draw (2.7,-1.3) node {$A^{\bar{z},d}(p_4)$};
\end{tikzpicture}
&= 
\langle \{ A^{z,a} (p_1), A^{z,b}(p_2) \}, \{ A^{z,c}(p_3), A^{\bar{z},d}(p_4) \} \rangle   \ ,
\end{align}
whose kinematic numerator evaluates to 
\begin{align}
\langle \{ A^{z,a} (p_1), A^{z,b}(p_2) \}, \{ A^{z,c}(p_3), A^{\bar{z},d}(p_4) \} \rangle  
& \sim (p_1-p_2)(2p_3 + p_4)~.
\end{align}
Note that the external polarization vectors are omitted, since they are just a multiplicative factor.
Summing over cyclic permutations of legs $(1,2,3)$ this numerator immediately vanishes. This numerator can equivalently be written in terms of the nested bracket as $\langle\big\{\{ A^{z,a}_1, A^{z,b}_2 \}, A^{z,c}_3\big\}, A^{\bar{z},d}_4\rangle$, for which the Jacobi identity holds. Let us explore this algebra in more detail. 
  On $\mathbb{R}^2$ we can introduce the basis for $\Omega^{(1,0)}$ and $\Omega^{(0,1)}$ differential forms
 \be
 L^{(1,0)}(p, \bar{p}) = e^{\frac{i}{2} (p\bar{z} + \bar{p}z)} dz~,~~~~ L^{(0,1)} (p, \bar{p}) =e^{\frac{i}{2} (p\bar{z} + \bar{p}z)} d\bar{z}~,
 \ee
  so that the brackets  (\ref{2D-bracket1}) and (\ref{2D-bracket2}) have the following structure constants
  \bea
  && \{ L^{(1,0)}(p_1, \bar{p}_1), L^{(1,0)}(p_2, \bar{p}_2) \} = \frac{i}{2} (p_2 - p_1)  \delta^2 (p_1+p_2 - p_3)  L^{(1,0)}(p_3, \bar{p}_3) \nn\\
    && \{ L^{(1,0)}(p_1, \bar{p}_1) , L^{(0,1)}(p_2, \bar{p}_2) \} = \frac{i}{2} (2p_1 + p_2) \delta^2 (p_1+p_2 - p_3) L^{(0,1)}(p_3, \bar{p}_3) \nn
  \eea
   and the pairing is defined by the only non-zero component as follows 
  \be
   \langle L^{(1,0)}(p_1, \bar{p}_1), L^{(0,1)}(p_2, \bar{p}_2) \rangle = - 2\pi i \delta^2 (p_1+p_2)~,
  \ee
   where $\delta^{2}$ is short-hand notation for the real two dimensional delta function. 
      Direct computation gives the following expression for the nested bracket,
      \begin{align}
    &   \langle \{ \{ L^{(1,0)}(p_1, \bar{p}_1), L^{(1,0)}(p_2, \bar{p}_2) \}, L^{(1,0)}(p_3, \bar{p}_3)\}, L^{(0,1)}(p_4, \bar{p}_4) \rangle \nn  \\
    & \hspace{2cm}=    \langle \{  L^{(1,0)}(p_1, \bar{p}_1), L^{(1,0)}(p_2, \bar{p}_2) \}, \{ L^{(1,0)}(p_3, \bar{p}_3), L^{(0,1)}(p_4,  \bar{p}_4) \rangle \nn \\
    & \hspace{2cm}=\frac{\pi i}{2} (p_2 - p_1) (2p_3 + p_4) \delta^{2} (p_1 + p_2 + p_3 + p_4)~. 
      \end{align}
   If we consider the sum of this expression with its cyclic permutation in labels $(1,2,3)$ we should get zero due to the Jacobi identity for the bracket.
 Indeed we find from the cyclic sum 
   \be
     (p_2 - p_1) (2p_3 + p_4) + (p_1 -p_3)(2p_2+p_4) + (p_3-p_2)(2p_1+p_4)=0~,
   \ee
    which is identically satisfied as expected.

For loops it is convenient to work in superspace, such that all ghosts and physical fields are handled together. For this we introduce the momentum-space representation of the differential operators
  $D$ and $D^\dagger$,
\begin{align}
   D &=  \frac{i}{2} \theta^z \bar{p} + \frac{i}{2} \theta^{\bar{z}} p +   \frac{i}{2}\frac{\partial}{\partial\zeta} \theta^z \theta^{\bar{z}} \\
   D^\dagger &= -2i p \frac{\partial}{\partial \theta^{z}}   + \Big(i\frac{\partial}{\partial\theta^{\bar{z}}} \theta^{\bar{z}} - 
   i\theta^{\bar{z}} \frac{\partial}{\partial\theta^{\bar{z}}} \Big)\frac{\partial}{\partial\zeta}~.
\end{align}
The superfield propagator is then defined by the following two conditions,
 \be
  D^\dagger  G(p, \bar{p}, \theta^z,  \theta^{\bar{z}}, \zeta ) = 0 
 \ee
 and 
  \be
  D   G(p, \bar{p},  \theta^z,  \theta^{\bar{z}}, \zeta ) = \theta^x \theta^y \zeta= \frac{i}{2} \theta^z \theta^{\bar{z}} \zeta ~, 
 \ee
  where on the right-hand side we have a delta function in the odd variables. 
 These conditions are solved by,
 \be\label{2Dsuperspaceprop}
  G(p, \bar{p},  \theta^z,  \theta^{\bar{z}}, \zeta ) = \frac{i}{2} \frac{D^\dagger ( \theta^z \theta^{\bar{z}} \zeta)}{p\bar{p}} 
  =
  \frac{\theta^{\bar{z}} \zeta p}{p\bar{p}} +   \frac{\theta^z \theta^{\bar{z}}}{2 p\bar{p}}\ ,
 \ee
  and thus the superfield correlator has the form
  \be
   \langle {\cal A}^a (p, \bar{p}, \theta^z,  \theta^{\bar{z}}, \zeta )  {\cal A}^b (0) \rangle  = G(p, \theta^z,  \theta^{\bar{z}}, \zeta ) \delta^{ab}~.
 \ee
The three-point vertex in superspace just involves integration over the odd coordinates,
\begin{eqnarray} \label{superfieldcubic}
\begin{tikzpicture}[baseline={(0, -0.1cm)}]
\draw[thick] (0,0.75) -- (0,0);
\draw[thick] (0,0) -- (0.87*0.75,-0.5*0.75);
\draw[thick] (0,0) -- (-0.87*0.75,-0.5*0.75);
\node at (0.05,-0.4) {${}_{(\theta,\zeta)}$};
\end{tikzpicture}
~~=~~
\int d^2 \theta d\zeta \ ,
\end{eqnarray}
and it is assumed that position space momentum is conserved in the vertex.
Using these superspace Feynman rules with external states $\mathcal{A}_i=\theta a_i + \bar{\theta} \bar{a}_i + 2\zeta p_i a_i$ we find the following numerator for the box diagram,
\begin{align}
\begin{tikzpicture}[baseline={(0, 0.2cm)},scale=0.75,every node/.style={scale=0.6}]
\draw[thick] (-1,0) -- (1,0);
\draw[thick] (-1,1) -- (1,1);
\draw[thick] (0.5,0) -- (0.5,1);
\draw[thick] (-0.5,0) -- (-0.5,1);
\node at (-1.25,0) {$1$};
\node at (-1.25,1) {$2$};
\node at (0,0.2) {$\ell$};
\node at (1.25,0) {$4$};
\node at (1.25,1) {$3$};
\end{tikzpicture} 
 ~=& ~ \int d^2 \theta d^2\tilde{\theta} d\zeta d \tilde{\zeta}\, \mathcal{A}_4 \delta_{\theta,\tilde{\theta}}^3
 ~ D^\dagger\Big(
 \mathcal{A}_3
 	D^\dagger\big(
 	\mathcal{A}_2
 		D^\dagger(
 			\mathcal{A}_{1}\tilde{D}^\dagger(\delta^3_{\theta,\tilde{\theta}})
 			)
 	\big)
 \Big) \nonumber \\
 =& ~
4 l^2 (p_1^2 + p_2^2 + p_2 p_3 + p_3^2 + p_1 (p_2 + p_3)) \nonumber \\
&+2 p_1 (p_1 + p_2 + p_3) (2 p_1^2 + p_1 (3 p_2 + p_3) + 
   p_2 (p_2 + 5 p_3)) \nonumber\\
&+
   2 l \big(4 p_1^3 + p_2 (p_2 + p_3) (2 p_2 + p_3)   \nonumber \\
   &\hspace{2cm}
   +p_1^2 (7 p_2 + 5 p_3) + 
   p_1 (5 p_2^2 + 6 p_2 p_3 + 3 p_3^2)\big)
\end{align}
where the loop momentum $l$ is assumed to flow clockwise, and momentum conservation $p_1+p_2+p_3+p_4=0$ was used with all momenta incoming. As explained earlier, only the incoming $a_i$ polarization factors play a role in this numerator, but they appear as a simple multiplicative factor $a_1 a_2 a_3 a_4$ which has been omitted. To check the Jacobi identity at one loop, we need also the triangle diagram whose numerator is,
\begin{align}
\begin{tikzpicture}[baseline={(0, 0.2cm)},scale=0.75,every node/.style={scale=0.6}]
\draw[thick] (-1,0) -- (1,0);
\draw[thick] (-1,1) -- (0,1);
\draw[thick] (0,1) -- (1,1);
\draw[thick] (0.5,0) -- (0,0.75);
\draw[thick] (-0.5,0) -- (0,0.75);
\draw[thick] (0,0.75) -- (0,1);
\node at (-1.25,0) {$1$};
\node at (-1.25,1) {$2$};
\node at (0,0.20) {$\ell$};
\node at (1.25,0) {$4$};
\node at (1.25,1) {$3$};
\end{tikzpicture} 
 ~&= ~ \int {d^2 \theta d^2\tilde{\theta} d\zeta d \tilde{\zeta} \, \mathcal{A}_4 \delta_{\theta,\tilde{\theta}}^3}~ D^\dagger\Big(
D^\dagger(\mathcal{A}_2\mathcal{A}_3) 
 {D^\dagger \big(\mathcal{A}_{1}\tilde{D}^\dagger(\delta^3_{\theta,\tilde{\theta}})\big)}\Big)
 \nonumber\\
 &= -2 (p_2 - p_3) \big(p_1 (p_1 + p_2 + p_3) (2 p_1 + p_2 + p_3) \nonumber \\
 &\hspace{3cm}+ 
   2 l (p_1^2 + p_1 (p_2 + p_3) + (p_2 + p_3)^2)\big) \ .
\end{align}
With these two numerators the Jacobi identity follows at one-loop level,
\begin{equation*}
\begin{tikzpicture}[baseline={(0, 0.2cm)},scale=0.75,every node/.style={scale=0.6}]
\draw[thick] (-1,0) -- (1,0);
\draw[thick] (-1,1) -- (1,1);
\draw[thick] (0.5,0) -- (0.5,1);
\draw[thick] (-0.5,0) -- (-0.5,1);
\node at (-1.25,0) {$1$};
\node at (-1.25,1) {$2$};
\node at (0,0.2) {$\ell$};
\node at (1.25,0) {$4$};
\node at (1.25,1) {$3$};
\end{tikzpicture} 
+
\begin{tikzpicture}[baseline={(0, 0.2cm)},scale=0.75,every node/.style={scale=0.6}]
\draw[thick] (-1,0) -- (1,0);
\draw[thick] (-1,1) -- (0,1);
\draw[thick] (0,1) -- (1,1);
\draw[thick] (0.5,0) -- (0,0.75);
\draw[thick] (-0.5,0) -- (0,0.75);
\draw[thick] (0,0.75) -- (0,1);
\node at (-1.25,0) {$1$};
\node at (-1.25,1) {$2$};
\node at (0,0.20) {$\ell$};
\node at (1.25,0) {$4$};
\node at (1.25,1) {$3$};
\end{tikzpicture} 
-
\begin{tikzpicture}[baseline={(0, 0.2cm)},scale=0.75,every node/.style={scale=0.6}]
\draw[thick] (-1,0) -- (1,0);
\draw[thick] (-1,1) -- (1,1);
\draw[thick] (0.5,0) -- (0.5,1);
\draw[thick] (-0.5,0) -- (-0.5,1);
\node at (-1.25,0) {$1$};
\node at (-1.25,1) {$3$};
\node at (0,0.2) {$\ell$};
\node at (1.25,0) {$4$};
\node at (1.25,1) {$2$};
\end{tikzpicture}
=
0 \ .
\end{equation*}
Of course, the same Jacobi identity could be extracted without evaluating the numerators and only using the second-order property of $D^\dagger$.

 \section{Summary}\label{s:summary}
 
 In this work we formalized and generalized observations from  \cite{Ben-Shahar:2021zww}  regarding the off-shell color-kinematics duality. 
  We have discussed  theories which can be recast in terms of formal Chern-Simons theory whose gauge fixing is encoded in an operator $D^\dagger$
   of degree $-1$. We have argued that if this operator satisfies three  properties listed in Section  \ref{s:formal}  then the gauge-fixed action can be written in the elegant form 
  $$S = \frac{1}{2} \langle {\cal A}^a, \square {\cal A}^b \rangle \delta_{ab} +  \frac{1}{6}
     \langle {\cal A}^a, \{ {\cal A}^b , {\cal A}^c\} \rangle  f_{abc}~.$$
 If $D^\dagger$ is a second-order operator then $\{~,~\}$ is a Lie bracket and $\langle~,~\rangle$ an invariant pairing. This algebraic description of 
  the gauge-fixed action is rather universal and true both for flat $\mathbb{R}^d$ and compact curved spaces. 
   In $\mathbb{R}^d$ these structures lead to off-shell color-kinematics duality.  Following ideas from  \cite{Ben-Shahar:2021zww}  it is straightforward to write down 
    the formal  double-copy theories for 4D BF-theory and 2D Yang-Mills, these taking the general form,
\begin{equation}\label{dcaction}
S = \int d\mu \, d\bar{\mu}\, \delta^d(x-\bar{x})\Big(
	\Phi \frac{D\bar{D}}{\Box}\Phi + \frac{1}{3!} \Phi^3
\Big) \ ,
\end{equation}  
where we introduced two copies of the fermionic coordinate spaces and we use an even degree superfield $\Phi$ which is expanded in both copies of the fermionic coordinates.
Similar double-copy actions have been constructed, involving Chern-Simons as well as (self-dual) Yang-Mills theory \cite{Ben-Shahar:2021zww,Bonezzi:2023pox,Borsten:2023paw,Bonezzi:2024dlv}, however in this context
we do not know if the actions \eqref{dcaction} correspond to some gauge-fixed version of a gravity theory. We hope to investigate this in future work. 

Yang-Mills theory in four and other dimensions can also be recast as a formal Chern-Simons theory.
As we will show in forthcoming paper  \cite{in-progress}, one can find the operator $D^\dagger$ which satisfies the required properties \eqref{basic-prop-1}-\eqref{basic-prop-3}, however, this operator is no longer necessarily of second order, therefore yielding other interesting algebraic structures in the gauge-fixed action.

\subsubsection*{Acknowledgements}

We are grateful to Henrik Johansson for helpful comments on an earlier draft.  Additionally, M.\ B.\ S. would like to thank Roberto Bonezzi for insightful discussions. The research of M.\ Z.\ 
   is  supported by the VR excellence center grant ``Geometry and Physics'' 2022-06593.
The research of M.\ B.\ S is funded by the European Union through the European Research Council under grant ERC Advanced Grant 101097219 (GraWFTy). Views and opinions expressed are however those of the authors only and do not necessarily reflect those of the European Union or European Research Council Executive Agency. Neither the European Union nor the granting authority can be held responsible for them.
\appendix
\section{Conventions on $\mathbb{R}^d$}\label{app-A}

In this Appendix we summarize the conventions we use for the differential forms and the Fourier transform on $\mathbb{R}^d$
 and some specific formulas for $\mathbb{R}^2$ in complex coordinates. 
 
 \subsection{the case of $\mathbb{R}^d$}

Consider the differential forms $\Omega^\bullet (\mathbb{R}^d)$ which can be identified with the functions on $C^\infty (T[1] \mathbb{R}^d)$
 with the Grassmann variables $\theta^\mu$.  On this space we define the de Rham operator and its adjoint as follows
\be
 d = \theta^\mu \frac{\partial}{\partial x^\mu}~,~~~~ d^\dagger = - \frac{\partial^2}{\partial \theta_\mu \partial x^\mu}~,
\ee
 where we have assumed the canonical flat metric on $\mathbb{R}^d$.  Thus the Laplace operator  in flat space would be 
 \be
   \{ d , d^\dagger \} = \square = - \frac{\partial}{\partial x^{\mu}}\frac{\partial}{\partial x_{\mu}}~. 
 \ee
 For the function $f \in C^\infty (T[1] \mathbb{R}^d)$ we define the  Fourier transform along even coordinates as follows 
\be\label{FT-conventions-gen}
  f(x, \theta) = \frac{1}{(2\pi)^{d/2}} \int d^d p~ e^{i p \cdot x} \hat{f}(p, \theta)
\ee
and its inverse as 
\be
  \hat{f}(p, \theta) = \frac{1}{(2\pi)^{d/2}} \int d^d x~ e^{-i p \cdot x} f(x, \theta)~.
\ee
   We use the following conventions for the delta function 
 \be
  \delta(x) = \frac{1}{(2\pi)^d} \int d^d p~ e^{i p \cdot x}~.
 \ee
 In momentum representation the relevant operators become
 \be
  d= i p_\mu \theta^\mu~,~~~~
  d^\dagger = - i p_\mu \frac{\partial }{\partial \theta_\mu}~,~~~~
  \square = p_\mu p^\mu \equiv p^2~.
 \ee
 
\subsection{the case of $\mathbb{R}^2$}\label{s:2dconventions}

Now let us specialize to the case of $\mathbb{R}^2$ with the coordinates $(x,y)$. $\mathbb{R}^2$ is equipped with the canonical flat 
 K\"ahler structure with the complex coordinates defined as 
 \be
  z= x+ i y~,~~~~~\bar{z}= x- i y \ ,
 \ee
  and for the odd coordinates
  \be
   \theta^z = \theta^x + i \theta^y~,~~~~~\theta^{\bar{z}} = \theta^x - i \theta^y~. 
  \ee
 The Dolbeault operators are defined as follows
\begin{equation}
\partial = \theta^z\partial_z \ , \hspace{2cm} \bar{\partial} = \theta^{\bar{z}}\partial_{\bar{z}}
\end{equation}
and their adjoint operators as
\begin{equation}
\partial^\dagger = -2\frac{\partial}{\partial\theta^z}\partial_{\bar{z}} \ , \hspace{2cm} \bar{\partial}^\dagger = -2\frac{\partial}{\partial\theta^{\bar{z}}}\partial_{{z}}~.
\end{equation}
Note that the Laplace operator is $\square = -\partial_x^2-\partial_y^2=-4\partial_z\partial_{\bar{z}}$  in complex
 coordinates. Here the volume form ${\rm vol} = dx \wedge dy$ coincides with the K\"ahler form $\omega$ which can be written in 
 the super-language as follows
\begin{equation}
\omega = \frac{i}{2}\theta^z\theta^{\bar{z}}
\end{equation}
 which together with the operation
\begin{equation}
\Lambda = 2i \frac{\partial}{\partial\theta^{{z}}}\frac{\partial}{\partial\theta^{\bar{z}}}
\end{equation}
 obey the standard K\"ahler identities.  Using the flat metric we define the Hodge star operation which can be rewritten 
  in terms of superfields as the following operation  
\begin{align}
\star &= \theta^x\theta^y + \theta^y\frac{\partial}{\partial\theta^x} - \theta^x\frac{\partial}{\partial\theta^y} + \frac{\partial}{\partial\theta^y}\frac{\partial}{\partial\theta^x} \\
&=\frac{i}{2}\theta^{z}\theta^{\bar{z}} - i(\theta^z\frac{\partial}{\partial\theta^z} - \theta^{\bar{z}}\frac{\partial}{\partial\theta^{\bar{z}}}) + 2i \frac{\partial}{\partial\theta^{{z}}}\frac{\partial}{\partial\theta^{\bar{z}}}~,
\end{align}
 where we have used the above conventions to rewrite in the complex coordinates. 
Thus within our conventions we have the following action on 1-forms
\be
  \star \omega^{(1,0)} = - i \omega^{(1,0)}~,~~~~~ \star \omega^{(0,1)} =  i \omega^{(0,1)}~.
\ee
 In our discussions the following operator plays an important role
\be
  {\cal I} = i\frac{\partial}{\partial\theta^{\bar{z}}} \theta^{\bar{z}} - i\theta^{\bar{z}} \frac{\partial}{\partial\theta^{\bar{z}}} ~.
\ee
 This operator ${\cal I}$  acts on 1-forms as $-\star$, on 0-forms as multiplication by $i$ and on 2-forms as multiplication by $-i$. The operator satisfies the following algebra
 \be
  {\cal I}^2 = -1~,~~~~~{\cal I}^{-1} = - {\cal I}
 \ee
  and moreover 
  \be
   \partial^\dagger {\cal I} = {\cal I} \partial^\dagger~.
  \ee
 Finally, rewriting (\ref{FT-conventions-gen}) on $\mathbb{R}^2$ in complex coordinates we get 
in 2D
\be
  f(z, \bar{z}, \theta) = \frac{1}{2\pi} \int d^2 p~ e^{\frac{i}{2}( \bar{p} z + p \bar{z})} \hat{f}(p, \bar{p}, \theta)
\ee
 where  $p = p_x + i p_y$.

\section{Brackets and operators}\label{app-B}

In this Appendix we  review  the algebraic relation between operators and the corresponding derived brackets. 
 The more general exposition of this subject can be found in  \cite{Voronov_2005}. 

 Consider a super commutative associative algebra ${\cal C}$ with unit $1$. We have an operator $D^\dagger : {\cal C} \rightarrow{\cal C}$.
  We define the following commutator 
  \be
   [D^\dagger, a]  b = D^\dagger (a b) - (-1)^{|a| {\rm deg} (D^\dagger)} a D^\dagger b~, 
  \ee
   acting on $ b \in {\cal C}$. An operator $D^\dagger$ is of $k$-order if the following nested commutator is zero
 \be
 [...   [ [D^\dagger, a_1] , a_2], ... , a_{k+1}] =0~.
 \ee
 For example, an operator $D^\dagger$ is of 1st order if the following holds
 \be
  [[D^\dagger, a_1], a_2]=0~,
 \ee
  which is equivalent to the well-known Leibniz identity for first order operators. 
  If an operator $D^\dagger$ is of second order then  the following holds
  \be
    [[[D^\dagger, a_1], a_2], a_3]=0~. 
  \ee
   We can define the brackets 
   \be
    \{ a_1, a_2 , ..., a_n \} \equiv  [...[[D^\dagger, a_1], a_2], ... , a_n ] (1)
   \ee
 where on the right hand side we act on the identity $1$. In \cite{Voronov_2005}  the main theorem states that if $(D^\dagger)^2=0$
  then the above brackets give rise to an $L_\infty$ algebra. 

In our case we require that $D^\dagger$ is an operator of degree $-1$ and $(D^\dagger)^2=0$. If $D^\dagger$ is an operator of degree $k$
 then the corresponding $L_\infty$ is truncated. In particular if $D^\dagger$ is of second order then the bracket
\be
\{a_1, a_2 \} = [ [D^\dagger, a_1] , a_2] (1) = D^\dagger (a_1 a_2) - (-1)^{|a_1|} a_1 (D^\dagger a_2) - (D^\dagger a_1) a_2
\ee
 is a super Lie bracket.  Here for the sake of clarity we assumed that $D^\dagger (1)=0$. The classical example 
  of this situation is when  the algebra ${\cal C} = \Omega^\bullet (\Sigma_d)$ is the algebra of differential forms with wedge product
   and $d^\dagger$ is the operator of second order as defined above. Therefore $d^\dagger$ defines a Lie bracket on the differential forms
    with shifted degree. 

\cleardoublepage

\bibliographystyle{JHEP}
\bibliography{ref}{}

\end{document}